\documentclass[letterpaper]{article}
\usepackage[utf8]{inputenc}
\usepackage{fullpage}
\usepackage{tcolorbox}

\usepackage[hidelinks]{hyperref}

\newcommand\arcsec{\mbox{$^{\prime\prime}$}}%
\def\farcm{%
 \mbox{.\kern -0.7ex\raisebox{.9ex}{\scriptsize$\prime$}}%
}%
\def\farcs{%
 \mbox{%
  \kern  0.13ex.%
  \kern -0.95ex\arcsec%
  \kern -0.1ex%
 }%
}%
\makeatletter
\newcommand\ion[2]{#1$\;$\protect \textsmaller{\rmfamily\@Roman{#2}}}%
\newcommand\smion[2]{#1$\;$\protect \footnotesize{\rmfamily\@Roman{#2}}\small}%
\newcommand\scion[2]{#1$\;$\protect \tiny{\rmfamily\@Roman{#2}}\scriptsize}%
\makeatother
\newcommand\nodata{ ~$\cdots$~ }%
\newcommand{\feh}{[Fe/H]}

\newcommand\aj{AJ}

\newcommand\araa{ARA\&A}
\newcommand\apj{ApJ}
\newcommand\apjl{ApJ}
\newcommand\apjs{ApJS}

\newcommand\aap{A\&A}

\newcommand\mnras{MNRAS}

\newcommand\prc{Phys.~Rev.~C}
\newcommand\prd{Phys.~Rev.~D}

\newcommand\prl{Phys.~Rev.~Lett.}
\newcommand\pasa{PASA}
\newcommand\pasp{PASP}
\newcommand\pasj{PASJ}

\newcommand\nat{Nature}

\newcommand\fcp{Fund.~Cosmic~Phys.}

\newcommand\nphysa{Nucl.~Phys.~A}

\usepackage{natbib}
\usepackage{color}
\usepackage{amsmath}
\usepackage{xspace}
\usepackage{graphicx}
\usepackage{amssymb}
\usepackage{xifthen}
\usepackage{hyperref}
\usepackage[normalem]{ulem}
\usepackage{hhline}
\usepackage{makecell}

\title{Observations of $R$-process Stars in the Milky Way and Dwarf Galaxies \\\vspace{0.3cm}
\large{to appear in \\
\textit{Handbook of Nuclear Physics} -- Part III \\
Section 15 ``Supernovae and Neutron Star Mergers"\\
by \textit{Springer Nature} 2022}
}
\author{}

\date{}

\begin{document}

\maketitle

\begin{center}
    
{\large Anna Frebel} \\
{\small Department of Physics and Kavli Institute for Astrophysics and Space Research, Massachusetts Institute of Technology, Cambridge, MA 02139, USA} \\ 
{\small Joint Institute for Nuclear Astrophysics--Center for Evolution of the Elements (JINA), East Lansing, MI 48824, USA} \\[12pt]
{\large and} \\[12pt]
{\large Alexander P. Ji} \\ 
{\small Department of Astronomy \& Astrophysics, University of Chicago, 5640 S Ellis Avenue, Chicago, IL 60637, USA} \\
{\small Kavli Institute for Cosmological Physics, University of Chicago, Chicago, IL 60637, USA} \\
{\small Joint Institute for Nuclear Astrophysics--Center for Evolution of the Elements (JINA), East Lansing, MI 48824, USA}
\end{center}

\begin{center}
\section*{Abstract} 

This chapter presents an overview of the recent progress on spectroscopic observations of metal-poor stars with $r$-process element signatures found in the Milky Way’s stellar halo and satellite dwarf galaxies.
Major empirical lessons related to the origins of the $r$-process are discussed, including the universality of the observed $r$-process pattern and deviations from universality among the light $r$-process elements and actinides. Different astrophysical sites of the $r$-process based on theoretical expectations are presented, including common and rare supernovae and neutron star mergers. A major distinguishing factor between $r$-process sites is their delay time distribution. The best constraints on the detailed $r$-process pattern come from Galactic halo $r$-process stars, but these cannot provide information on the environment of the stars' birth gas clouds. Studying $r$-process enrichment within dwarf galaxies can remedy the situation despite the fact that high-resolution spectroscopic observations of individual stars in these systems are very difficult to obtain. A general overview of dwarf galaxy properties and chemical evolution expectations depending on their mass and star formation duration is provided. The $r$-process trends depend on the stellar mass and star formation durations of dwarf galaxies in a way that clearly shows that the $r$-process is rare, prolific, and has both prompt and delayed sources. This work complements ongoing theoretical heavy-element nucleosynthesis explorations and experimental measurements of the properties of $r$-process nuclei, such as with the Facility for Rare Isotope Beams.

\end{center}


\newpage
\section{Introduction}

Astronomical observations of cosmic objects provide insight into a large range of physics questions. One central question is that of the origin of the chemical elements through a variety of nucleosynthesis processes that occur in stars, supernova explosions, merging neutron stars and other objects. 

\begin{figure}[!b]
\centering
    \includegraphics[width=0.8\linewidth]{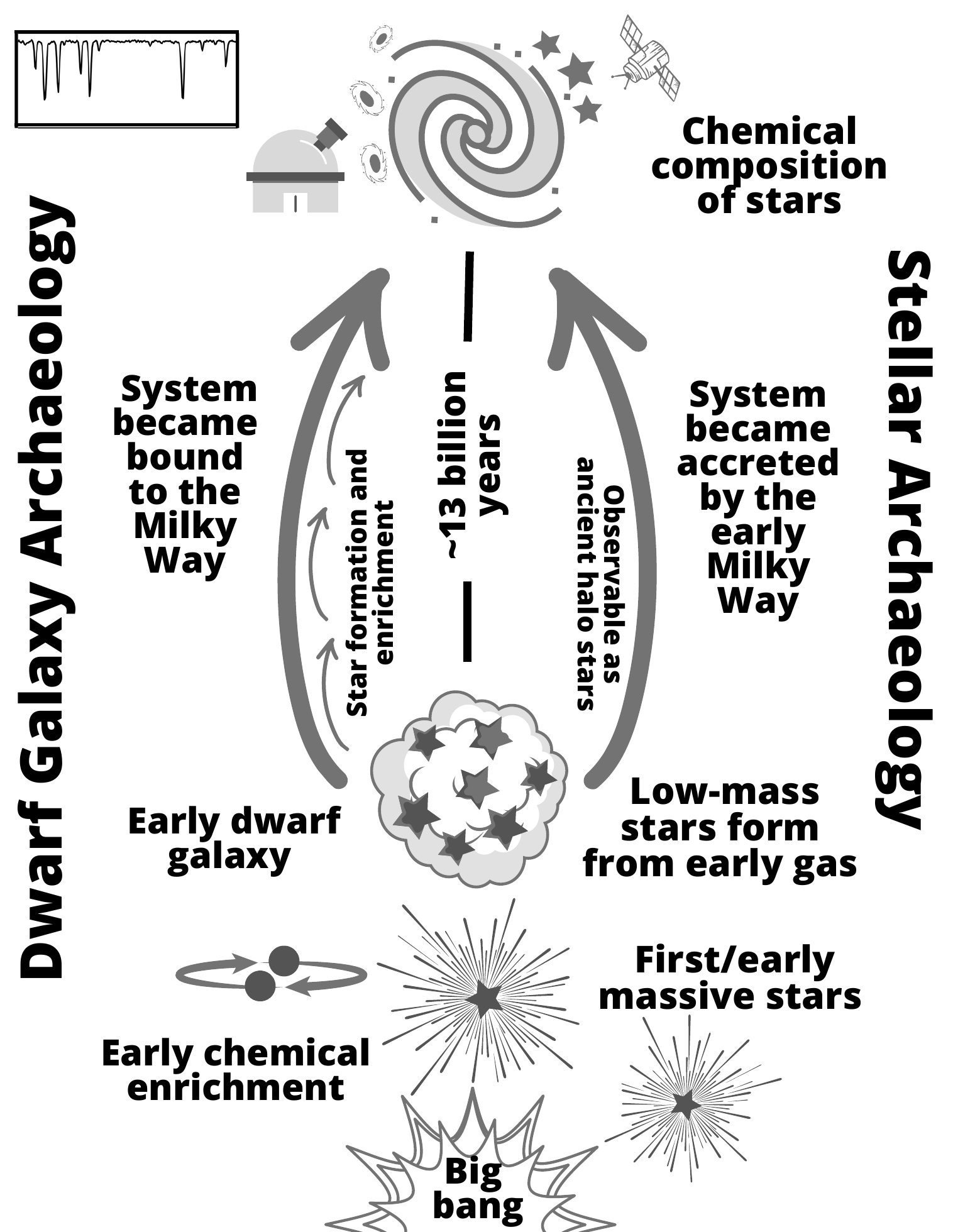}
    \caption{Overview of the concepts of stellar and dwarf galaxy archaeology. In both cases, observations of ancient stars in the Milky Way's halo as well as its orbiting dwarf satellite galaxies enable the reconstruction of the physical and chemical conditions of stellar birth gas clouds present within the earliest galactic systems that formed soon after the Big Bang as well as over all of cosmic history. Comparisons with theoretical models allows testing for specific sources and sites of nucleosynthesis and their effects on their environment.
    }
    \label{fig:overview}
\end{figure}

This chapter presents an overview of recent astronomical observations of heavy elements associated with $r$-process nucleosynthesis. The $r$-process (short for ``rapid neutron-capture process'') is responsible for synthesizing about half of the isotopes of elements heavier than iron in the solar system. The $s$-process (short for ``slow neutron-capture process'') is responsible for the other half.
The elements synthesized by these and other nucleosynthetic processes can be probed by studying the chemical composition of ancient, so-called ``metal-poor" stars. The elements preserved in these objects over billions of years enable us to study abundance signatures that arose from just a few (or even one) nucleosynthesis events that occurred prior to the star's birth in the early universe. This concept is often referred to as Stellar Archaeology, and illustrated in Figure~\ref{fig:overview}.
Indeed, having the means to observationally isolate a variety of clean nucleosynthesis signatures, and interpreting them within the framework of nuclear astrophysics, is vital for our understanding where and how the chemical elements were synthesized \citep{Frebel2015}.

In this context, it is important to begin with a definition of metal-poor stars. ``Metal" refers in astronomy to all elements heavier than H and He. ``Metal-poor" refers to an amount of all heavy elements that is less than that of the Sun, our reference star. Effectively that means, that the lower the overall metallicity, the earlier the star must have formed due to ongoing chemical enrichment events that only ever add new metals to the universe. With spectroscopic observations, the chemical composition of stars can be obtained to determine metallicity and abundances of individual elements. In most cases, the iron abundance of the star (given as ``[Fe/H]" in the bracket notation) will be used to represent its overall metallicity, and used interchangeably. Stellar chemical abundances of any two elements are then defined relative to the respective abundances in the Sun (e.g., \citealt{Asplund2009}) as \mbox{[A/B]}$
= \log(N_{\rm A}/N_{\rm B}) - \log(N_{\rm A}/N_{\rm B})_\odot$ where
$N_{\rm {A}}$ ($N_{\rm {B}}$) is the number of atoms of element A (B). Consequently, the Sun will have a metallicity of $\mbox{[Fe/H]}=0.0$. Stars with $\mbox{[Fe/H]}<-1.0$ are called ``metal-poor" stars. Stars with $\mbox{[Fe/H]}<-2.0$ are called ``very metal-poor" stars. An ``extremely metal-poor" star would have $\mbox{[Fe/H]}<-3.0$, which corresponds to one thousandth of the solar iron abundance. For better or for worse, the words ``very'' and ``extremely'' have a specific technical meaning when studying metal-poor stars. For a full classification, see table 1 in \citealt{Frebel2018}.

For nearly three decades by now, stellar archaeology has been carried out with a group of rare, old, very and extremely metal-poor stars that display large relative enhancements in heavy elements associated with the $r$-process. They are found in the outskirts of the Milky Way, in the Galactic halo \citep{Beers05, Sneden2008, Frebel2018}. 
Astronomers distinguish different $r$-process enhancement levels of these stars in the following way. First, $r$-process stars need to fulfill the criterion of $\mbox{[Ba/Eu]} < 0.0$ to ensure their neutron-capture element content does not originate with the $s$-process. Mildly enhanced $r$-process stars are defined to have $0.3 \le \mbox{[Eu/Fe]} \le 0.7$ and are termed $r$-I stars. Strongly enhanced $r$-process stars have $\mbox{[Eu/Fe]} > 0.7$; they are the $r$-II stars. An earlier definition had the transition from $r$-I to $r$-II at $\mbox{[Eu/Fe]} =1.0$ but recent works shows a naturally occurring break at 0.7 in the distribution, with about 2/3 of the $r$-process stars having values of $\mbox{[Eu/Fe]}<0.7$. Finally, the recent discovery of a star with $\mbox{[Eu/Fe]} = 2.3$ \citep{cain20} led to the introduction of $r$-III stars with exceedingly large Eu abundances of $\mbox{[Eu/Fe]} > +2.0$.

Metal-poor halo stars are thought to originate in small dwarf galaxies that have been tidally disrupted by the Milky Way \citep{Frebel2015}.
It was mostly impossible to associate these halo stars with their birth environments, but there are also intact dwarf satellite galaxies orbiting the Milky Way at larger distances.
Starting in the 2000s \citep[e.g.,][]{Shetrone01,shetrone03,Koch08} the advent of 6-10m diameter telescopes enabled direct study of the chemical abundances of stars in these dwarf galaxies (see reviews by \citealt{Venn04,Tolstoy2009,Frebel2015,Simon2019}).
At first, very few stars were discovered with clear $r$-process-enhanced nucleosynthetic signature outside the stellar halo (in the dwarf galaxy Ursa Minor and the globular cluster M15; \citealt{Otsuki06, aoki2007_cos82, Cohen2009}).

It thus came as a surprise in 2016 when the dwarf galaxy Reticulum\,II \citep{Ji2016b, Roederer2016} turned out to be an entire ``$r$-process galaxy" when seven of nine stars observed showed the $r$-process pattern and enhancement levels among the highest ever found. The existence of such stars in a known and quantifiable dwarf galaxy environment provided a novel constraint on the astrophysical production site of the $r$-process. Based on an estimate for the mass of Reticulum\,II and together with the enhancement level observed in the stars, it was concluded that likely a neutron star merger had occurred to provide copious amounts of $r$-process elements in this galaxy at very early times. Using entire dwarf galaxies as probes of chemical enrichment events -- and more broadly to constrain the physical and chemical conditions of early galaxies --  significantly builds on Stellar Archaeology, since the birth environment of dwarf galaxy stars can be estimated and taken into account. Accordingly, this is referred to as ``dwarf galaxy archaeology" and also shown in Figure~\ref{fig:overview}.



\begin{figure}
\centering
    \includegraphics[width=0.7\linewidth]{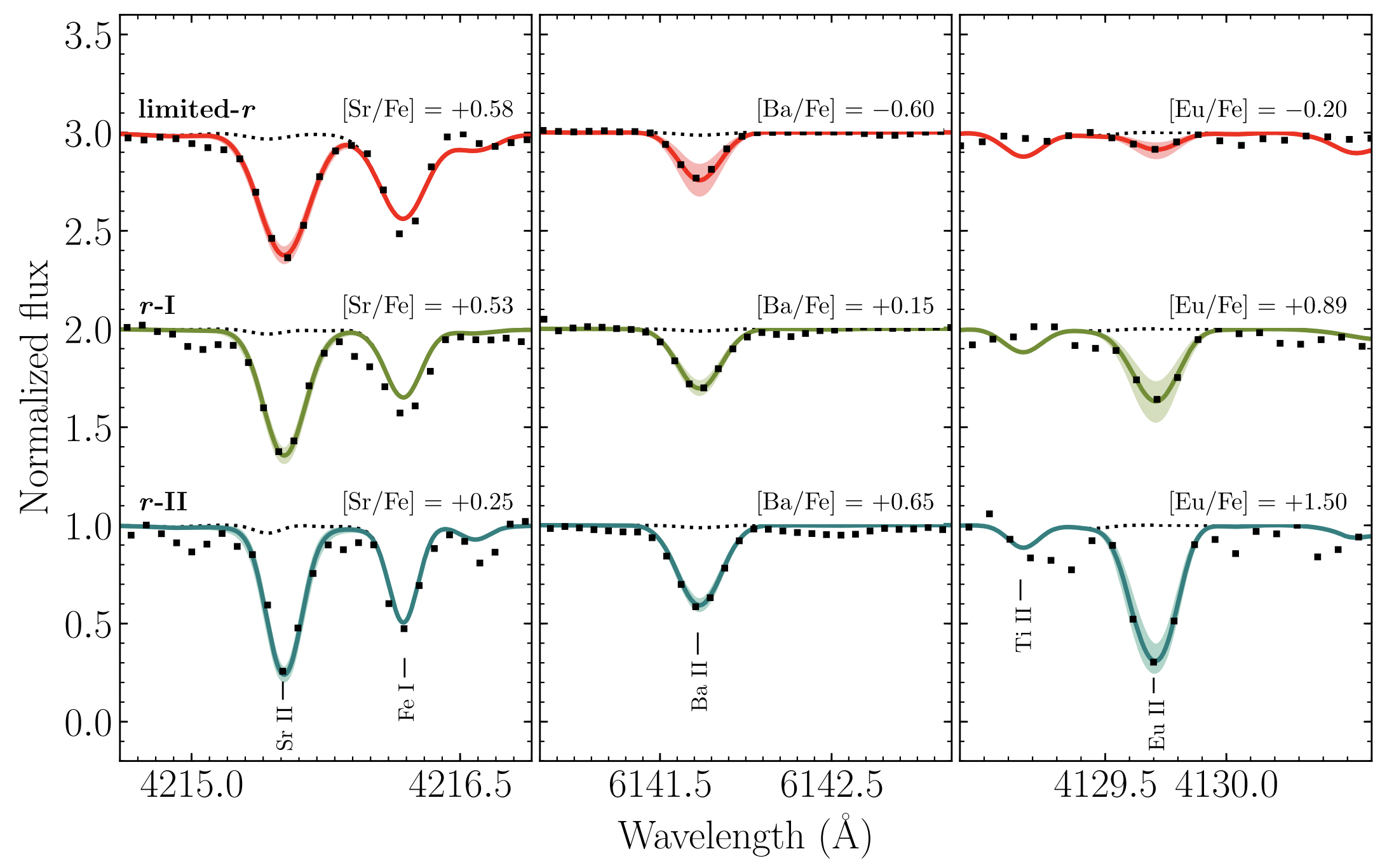} \\
    \includegraphics[width=0.7\linewidth]{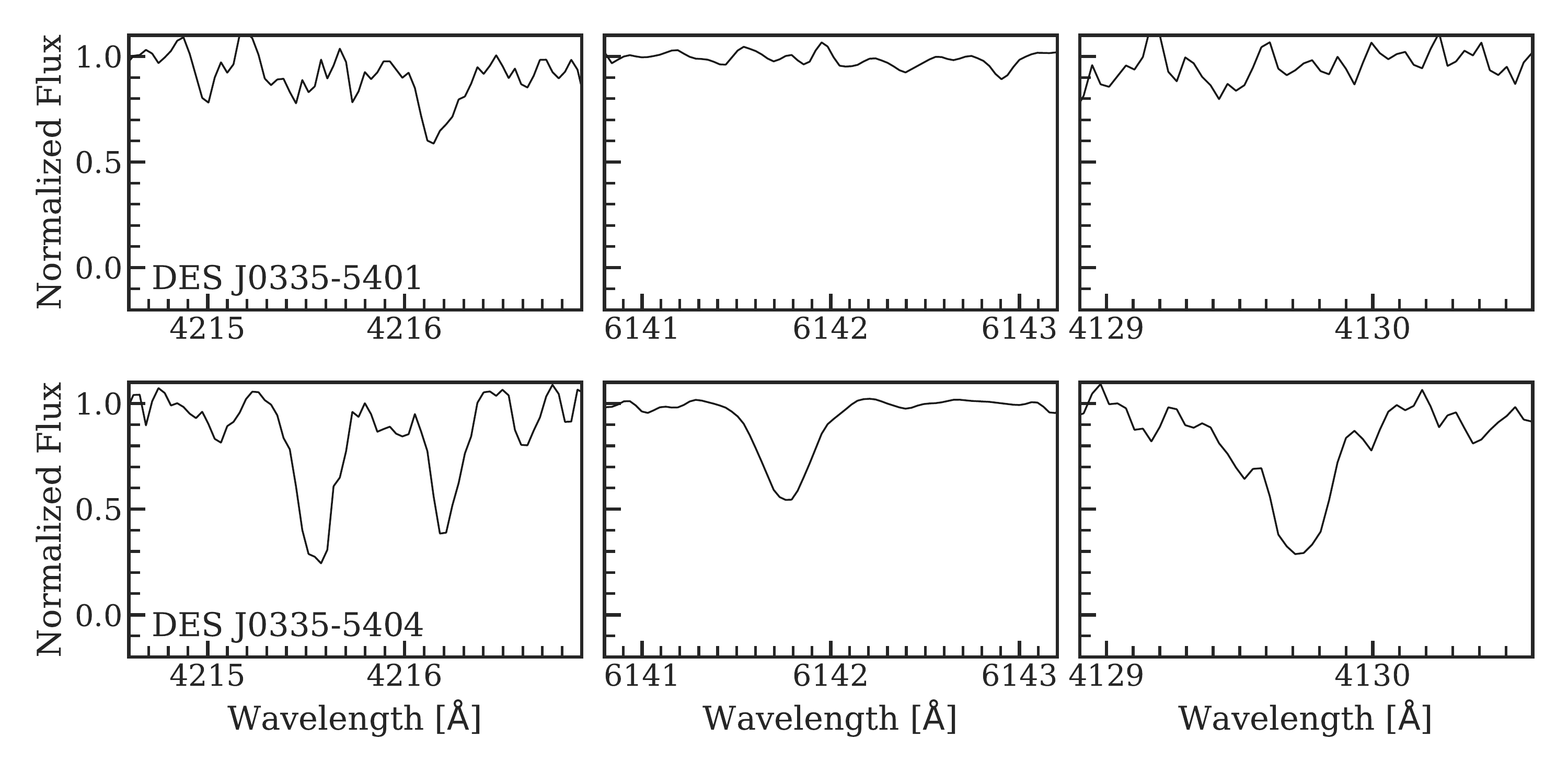}
    \caption{Portion of spectra of stars with different levels of neutron-capture element enhancement around the strong Sr, Ba, and Eu absorption lines at 4215\,{\AA}, 6141\,{\AA}, and 4129\,{\AA}, respectively. Top panels: A limited-$r$ halo star (top, J1812$-$4934), a mildly enhanced $r$-process halo star (middle, J1558$-$1224) stars, and a strongly $-r$-process enhanced halo star (bottom, J0246$-$1518). Figure taken from \citet{Hansen2018}. 
    Bottom panels:
    Two stars from the ultra-faint dwarf galaxy Reticulum\,II \citep{Ji2016b}. The top star is a typical ultra-faint dwarf galaxy star completely lacking neutron-capture element lines, while the bottom star is a strongly $r$-process enhanced star. The halo stars were observed with a relatively small 2.5m diameter telescope in 10-30 min, while the dwarf galaxy stars were observed with a 6.5m diameter telescope for 1-3 hours. Despite the vast difference in telescope diameter and exposure time, note the stark difference in signal-to-noise ratio of the spectra between the halo stars and Reticulum\,II stars.
    }
    \label{fig:spectra}
\end{figure}

Metal-poor halo stars and dwarf galaxies thus provide complementary ways to observationally access information about the origin of the $r$-process elements.
Metal-poor halo stars are relatively nearby ($<5$\,kpc), so it is possible to directly study the detailed composition of $r$-process nucleosynthesis across all the $r$-process peaks. This has led to many exquisite results and findings, e.g., the universality of the $r$-process, uranium measurements, actinide boost, second $r$-process peak abundances, that have driven this line of observational exploration of $r$-process nucleosynthesis. 
Dwarf galaxies are distant (30-200\,kpc) making it challenging to obtain stellar spectra from which to determine full $r$-process element patterns (see Figure~\ref{fig:spectra}), but the known birth environment of the stars provides the opportunity to understand the buildup of $r$-process material over time. The many dwarf galaxies sample a wide range of formation histories, providing a way to test the relative contributions of $r$-process sites with different rates, yields, and delay time distributions.
Ultimately, the goal is a unified picture of the origin of $r$-process elements in the intact dwarf galaxies, the tidally disrupted dwarf galaxies, and the Milky Way Galaxy itself. This approach highlights the importance of nuclear astrophysics that brings together results from astrophysics and nuclear physics in the quest to understand the $r$-process. The work with $r$-process stars and dwarf galaxies will thus provide information complementary to results to be obtained from experimental facilities such as the Facility for Rare Isotope Beams (FRIB).

The rest of this Chapter first describes lessons about $r$-process nucleosynthesis from Stellar Archaeology (Section~\ref{sec:halo}) with metal-poor halo stars. The next section briefly discusses different $r$-process sites and how a primary distinguishing factor is not which elements are made, but \textit{when} they are made (Section~\ref{candidates}). This motivates studying dwarf galaxies to understand the $r$-process. General information about dwarf galaxies is reviewed in Section~\ref{sec:dwarfgals}, and a discussion of lessons regarding the $r$-process origin from dwarf galaxies is in Section~\ref{sec:dwarfrproc}. A summary and the path ahead is discussed in Section~\ref{sec:conclusion}.

\section{$R$-process Nucleosynthesis in the Cosmos}\label{sec:halo}

Starting in the mid 1990's, astronomers began to occasionally find metal-poor stars that would show an unusual, strong enhancement in $r$-process elements. They would be located in the so-called halo of the Milky Way, depicting the extensive spherical region that envelopes the Galaxy's disk. The first such star (with $\mbox{[Fe/H]}\sim-3.1$) was CS22892-052 \citep{sneden1994,Sneden03}) which eventually became one of the most well-studied metal-poor stars, certainly in terms of number of elements. An impressive 57 elements were measured based on high-resolution spectroscopic data taken with ground and space based telescopes. Soon after, CS31082-001 was found \citep{Cayrel01, Hill02}, another strongly $r$-process enhanced star which now also enables Th \textit{and} U measurement. However, this star appeared to be enhanced in Th and U relative to other neutron-capture element abundances which lead to negative values when attempting to age date this star (although the Th/U ratio did deliver an age of $\sim$14 billion years). The actinide boost phenomenon was found, leading to even more questions about $r$-process nucleosynthesis, its astrophysical site, and frequency of occurrence.
All these initial observational findings of metal-poor halo stars enhanced in $r$-process elements showed without a doubt \textit{that} the $r$-process operated in the early universe, and that clean nucleosynthesis signatures could be extracted from stars through detailed chemical abundance analyses. It opened up a new branch of nuclear astrophysics by enabling the collection of the observable fingerprints of this process.

\subsection{Discovering $R$-process Stars in the Milky Way Halo and their Abundance Signatures}

Astronomers began to systematically search for more metal-poor $r$-process stars since it had become clear that they are out there -- just not in large numbers. 
In the early 2000's, an extensive observational campaign, the HERES survey \citep{Christlieb04, Barklem2005}, was started that would collect hundreds of so-called snapshot spectra of metal-poor red giants allowing sifting through many stars to check for the presence of the strong Eu absorption line at 4129\,{\AA}, the most prominent indicator for $r$-process enhancement. Absorption line strength is temperature dependent where, at fixed abundance, the cooler red giants will display a stronger, more easily to detect line than warmer main-sequence star counterparts. Since 2017, the $R$-Process Alliance (e.g., \citealt{sakari18,hansen18,ezzeddine20, holmbeck20}) has continued this work by conducting a multi-year observational program with telescopes in both the Northern and Southern hemisphere, with the goal of quadrupling the number of existing $r$-process enhanced stars. Thus far, more than 2000 candidates have been observed, of which $\sim$70 are strongly enhanced $r$-process stars with $\mbox{[Eu/Fe]}>0.7$ discovered by the $R$-Process Alliance. 

Detailed chemical abundance studies have by now been carried out for dozens of $r$-process stars based on high-resolution, high $S/N$ spectra obtained with telescopes such as the 8 meter ESO/VLT, 6.5 meter Magellan, and 8 meter Subaru \citep{Christlieb04, Honda05, holmbeck18, placco2020, cain20}. A handful have even been studied with UV spectra obtained with the Hubble Space Telescope and its STIS and COS spectrographs to obtain measurements of additional elements with transition only available in the UV spectral range \citep{Roederer2012, roederer2014phos, Roederer18_hd}. The challenge with UV measurements is that stars must be rather bright, fairly warm with respect to its surface temperature as to produce a larger relative UV flux, and of course highly enhanced in $r$-process elements.

\subsection{How Much $R$-process Material is Found in Old Halo Stars?}

\begin{figure*}[!h] 
\begin{center}
  \includegraphics[width=\linewidth]{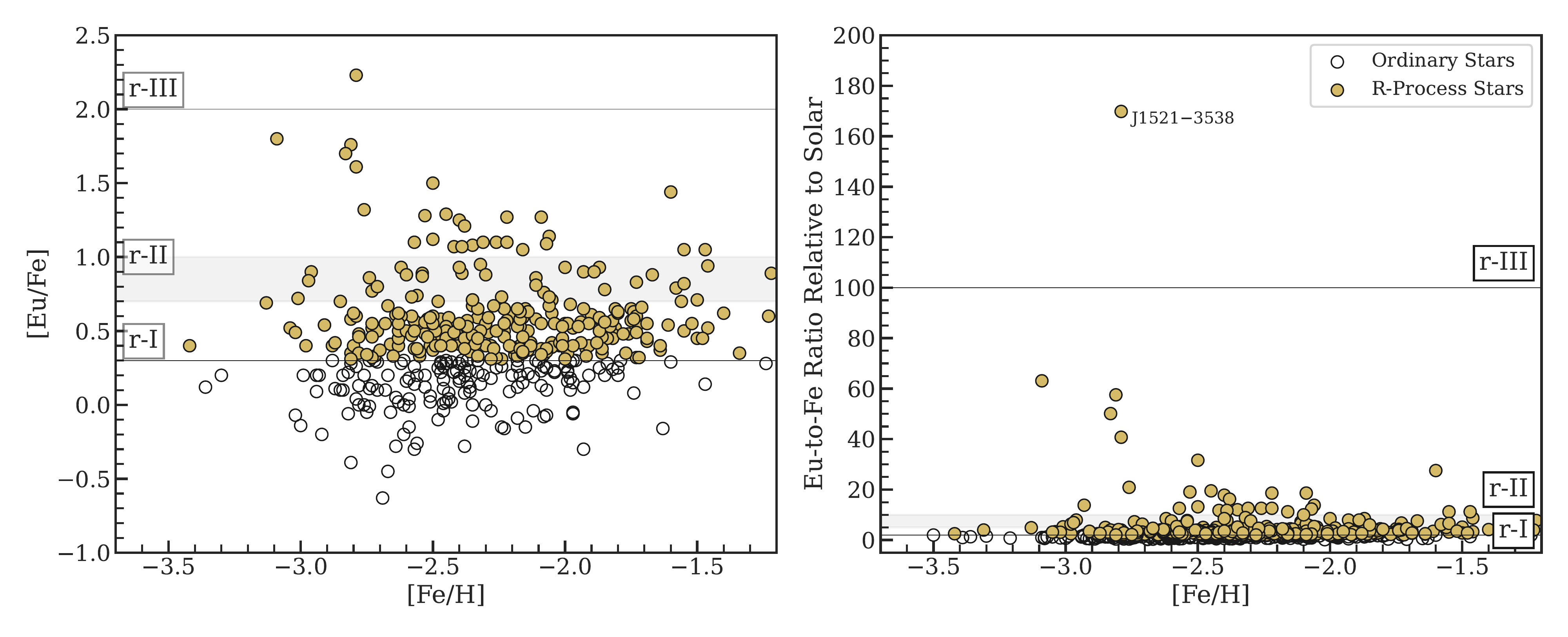}
   \caption{\mbox{Logarithmic [Eu/Fe]} and linear Eu-to-Fe ratio (relative to that of the Sun) plotted against \feh\ for various $r$-process enhanced observed by the $R$-Process Alliance. The adopted boundaries for $r$-I, $r$-II, and $r$-III stars are shown at $\mbox{[Eu/Fe]} = 0.7$ and 2.0. 
   For educational purposes, the older $r$-I-$r$-II boundary of $\mbox{[Eu/Fe]} = 1.0$ that was based on a smaller dataset is also shown.
   Data are from \citep{hansen18,sakari18,ezzeddine20, holmbeck20}. 
   The $r$-III star J1521$-$3538 is marked separately \citep{cain20}.
  Additional $r$-II stars can be found in JINAbase \citep{Abohalima18}. \label{fig:eufe}
   }
\end{center}
\end{figure*}

As part of the discovery process, candidate stars are identified by the Eu line to obtain the stars' Eu-to-Fe ratio. Moderate enhancement, termed $r$-I, has been originally defined as $0.3<\mbox{[Eu/Fe]}<1.0$ \citep{Beers05} but more recently been changed to $0.3<\mbox{[Eu/Fe]}<0.7$ \citep{holmbeck20} to better reflect a naturally occurring break in the abundance distribution. Regardless, $r$-I stars show ratios at least at a factor of 2 or higher than the Eu/Fe ratio of the Sun. Strong enhancement, termed $r$-II, now defined as $\mbox{[Eu/Fe]}>0.7$, a factor of at least 5 times higher compared to the Sun's ratio. By now, stars with much higher ratios has been found, up to $\mbox{[Eu/Fe]}\sim2.3$. Stars with $\mbox{[Eu/Fe]}>2.0$ have been termed $r$-III stars \citep{cain20, ezzeddine20}.  This is illustrated in Figure~\ref{fig:eufe}. These $r$-III stars display more than a factor of 100 higher Eu/Fe ratios than the Sun, providing evidence of significant contributions of Eu to the stars' early birth gas clouds, in relation to Fe contributions by other sources and as compared to the pre-solar nebula's enrichment history. The $r$-III stars are especially interesting for setting constraints on the maximum yield of potential $r$-process sites \citep[e.g.,][]{Macias2018,Macias2019,Siegel2021}

In this context, it is also interesting to look at the ``raw" abundance, log $\epsilon(Eu)$ or [Eu/H] for a more direct comparison to the amount of Eu present in the Sun (formed $\sim$9.2 billion years after the Big Bang). 
At $\mbox{[Fe/H]}\sim-3.1$ and with $\mbox{[Eu/Fe]}\sim1.5$ \citep{sneden2000}, CS22892-052 has still contains a significantly lower total Eu amount than the Sun of $\mbox{[Eu/H]}\sim-1.6$, or a factor of 40 less than the Sun's value. Still, CS22892-052 is called a strongly enhanced $r$-process star because the Eu amount relative to Fe is so unusual given it's (supposed) early formation time as indicated by the low [Fe/H] abundance (probably within the first billion years or so). Compare this to another star. At $\mbox{[Fe/H]}\sim-1.5$ and with $\mbox{[Eu/Fe]}\sim1.3$ \citep{Roederer18_hd}, the $r$-II star HD\,222925 has $\mbox{[Eu/H]}\sim-0.2$. This  is very close to the amount of Eu found in the Sun, and in fact the largest Eu amount discovered in any of the $r$-process stars. Models incorporating $r$-process nucleosynthesis, chemical evolution and star formation will need  to reproduce not just the observed $r$-process abundance signature of these stars but also explain their total amounts of Eu and other heavy elements.

\subsection{Frequency of $R$-process Stars in the Galactic Halo}

While the Galactic halo contains relatively many ancient metal-poor stars, the bulk of objects are much younger and metal-rich. This leaves metal-poor stars to be rare objects that are difficult to identify \citep{Beers05, Frebel2015}. 
The $r$-process enhanced stars are exceedingly rare and determining the occurrence rate has not been easy, especially when samples were still small. For a long time, for stars with $\mbox{[Eu/Fe]}>1.0$, the rate was about 5\% among metal-poor stars with $\mbox{[Fe/H]}<-2.0$ \citep{Barklem2005}. As of Spring 2022, hundreds of $r$-process are now known spanning a large range of enhancement levels. Specifically, $r$-II stars with $\mbox{[Eu/Fe]}>1.0$ found by the $R$-Process Alliance remain at 5\% (27 stars), but numbers increase to 12\% (70 stars) when taking $\mbox{[Eu/Fe]}>0.7$ \citep{Hansen2018,sakari18,ezzeddine20,holmbeck20}. According to the JINAbase database  \citep{Abohalima18} for metal-poor stars, 30 additional stars with $\mbox{[Eu/Fe]}>1.0$ are known (3\%) and 65 stars with $\mbox{[Eu/Fe]}>0.7$ (7\%), all for $\mbox{[Fe/H]}<-1.5$. This brings the total of known $r$-process halo metal-poor stars with  $\mbox{[Eu/Fe]}>1.0$ to 57, and 135 with $\mbox{[Eu/Fe]}>0.7$. Stars with Eu enhancement of $0.3<\mbox{[Eu/Fe]}<0.7$ are much more numerous, about 2-3 times as many of the strongly enhanced $r$-process stars. Interestingly, no metallicity dependence has been observed on these fractions \citep{holmbeck20}, although it should be noted that hardly any of these stars have $\mbox{[Fe/H]}<-3.3$.

\subsection{The Universality of the $R$-process between Barium and Platinum}

\begin{figure}
\centering
    \includegraphics[width=0.7\linewidth]{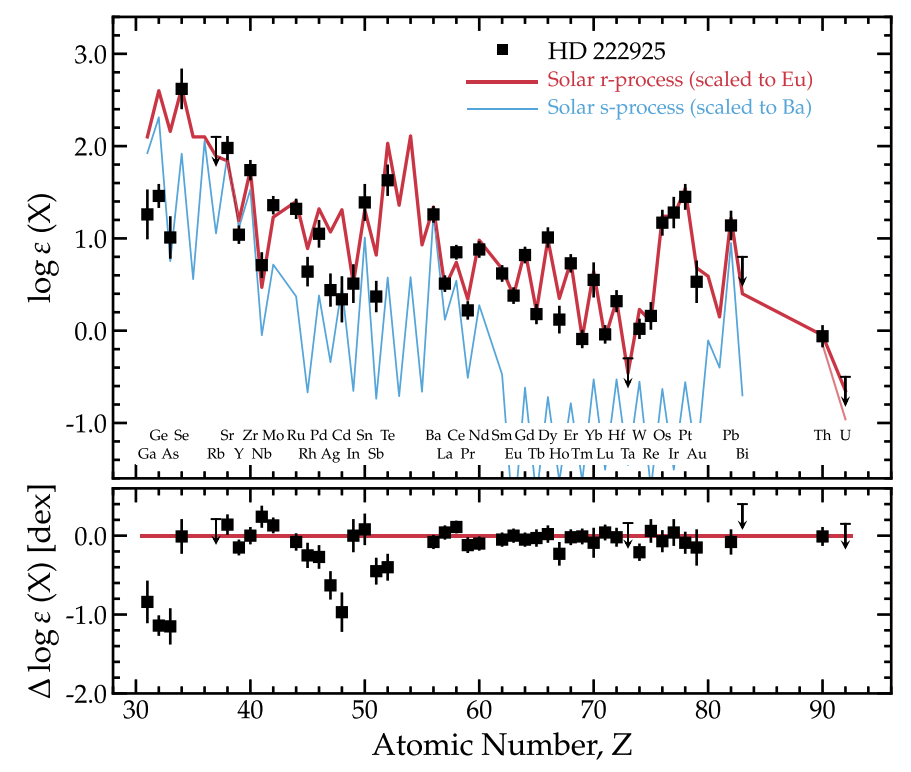}
    \caption{$R$-process pattern of the metal-poor red giant star HD\,222925 which has the most complete elemental abundance pattern of directly detected 42 $r$-process elements, based on ground-based and space-based spectroscopic data. The three elemental $r$-process peaks at $Z \sim 34$, $54$, and $78$ correspond to the isotopic abundance peaks at $A \sim 80$, $130$, and $195$ that originate from the closed neutron shells at $N=28$, $50$, and $82$. The match to the scaled solar $r$-process pattern is striking in the region from Ba to Pt. HD\,222925 does not appear to be an actinide boost star. Differences, however, are found for several trans-Fe elements (Ga, Ge, As) and among the lighter neutron-capture elements close to the second peak. 
    Figure taken from \citet{Roederer2022}.}
    \label{fig:rprocpattern}
\end{figure}

Ever since the discovery of the first $r$-process star, CS22892-052, it has been shown repeatedly that the derived abundance pattern of heavy neutron-capture elements in all these stars shows a universal behavior; that is the relative $\log \epsilon$(X) abundances (defined as $\log_{10}\epsilon$(X) = $\log_{10}$($N_{\rm X}$/N$_{\rm H}$) + 12 $= \mbox{[X/H]} + \log\epsilon_\odot$(X)) for elements barium and above always show the exact same distinct pattern. This is illustrated in Figure~\ref{fig:rprocpattern} for the metal-poor $r$-process enhanced star HD\,222925 \citep{Roederer2022}. For example, two stars with different metallicities would not share the same level of enrichment (higher metallicity stars tend to have higher overall abundances including those of neutron-capture abundances) but they would share the same \textit{relative} abundance pattern. This allows for using e.g., the [Ba/Eu] ratio as an indicator for observationally identifying $r$-process stars given that a value of $\mbox{[Ba/Eu]} =-0.8$ simply refers to that underlying pattern \citep{Sneden2008}.

This universality of the heavy $r$-process elements, in fact, even extends to the $r$-process residuals of the Sun -- stars covering a range of metallicities from $\mbox{[Fe/H]}\sim-1$ to $-3$ and the Sun show the same relative abundance pattern despite having vastly different overall levels of heavy element abundances. Note that the Sun's $r$-process residual is obtained by subtracting a theoretically derived $s$-process component from the Sun's total chemical abundance pattern \citep[e.g.,][]{Goriely1999,Sneden2008,Bisterzo2014,Prantzos2020}. This is necessary because the Sun formed from gas that experienced a variety of nucleosynthesis events over billions of years, including $s$- and $r$-process. As such, it is important to remember that the Sun's pattern is not a measured quantity as it is the case for the $r$-process abundances patterns obtained from metal-poor stars. Still, the universality seems to hold. Overall, the universality implies that no matter where and when in the universe, the $r$-process production of heavy elements of barium to the third peak elements  appears to remains the same irrespective of any astrophysical effects.

While the Sun's $r$-process residual pattern is a derived quantity, it uniquely has isotope abundances available in addition to element abundances. As such, it has been the long-standing benchmark star for comparisons with essentially all observational and theoretical results.  However, recently, the bright star HD\,222925 was discovered to be $r$-process-enhanced. Using both ground and spaced based observations it was possible to measure chemical abundances of 42 heavy neutron-capture elements from 31 $\le$ Z $\le$ 90 \citep{Roederer2022}. This makes HD\,222925 the star with the most elements measured besides the Sun. Going forward, HD\,222925 could be used as the benchmark for all further comparisons related to testing the site of the $r$-process, especially given the large deviations in the first $r$-process peak.
Table~\ref{tab:solarrproc} provides multiple different determinations of the solar $r$-process residual as well as the measurements of HD\,222925 \citep{Prantzos2020,Roederer2022}.

\subsection{Variations of Light Neutron-Capture Element Abundance}

\begin{figure}
\centering
    \includegraphics[width=0.9\linewidth]{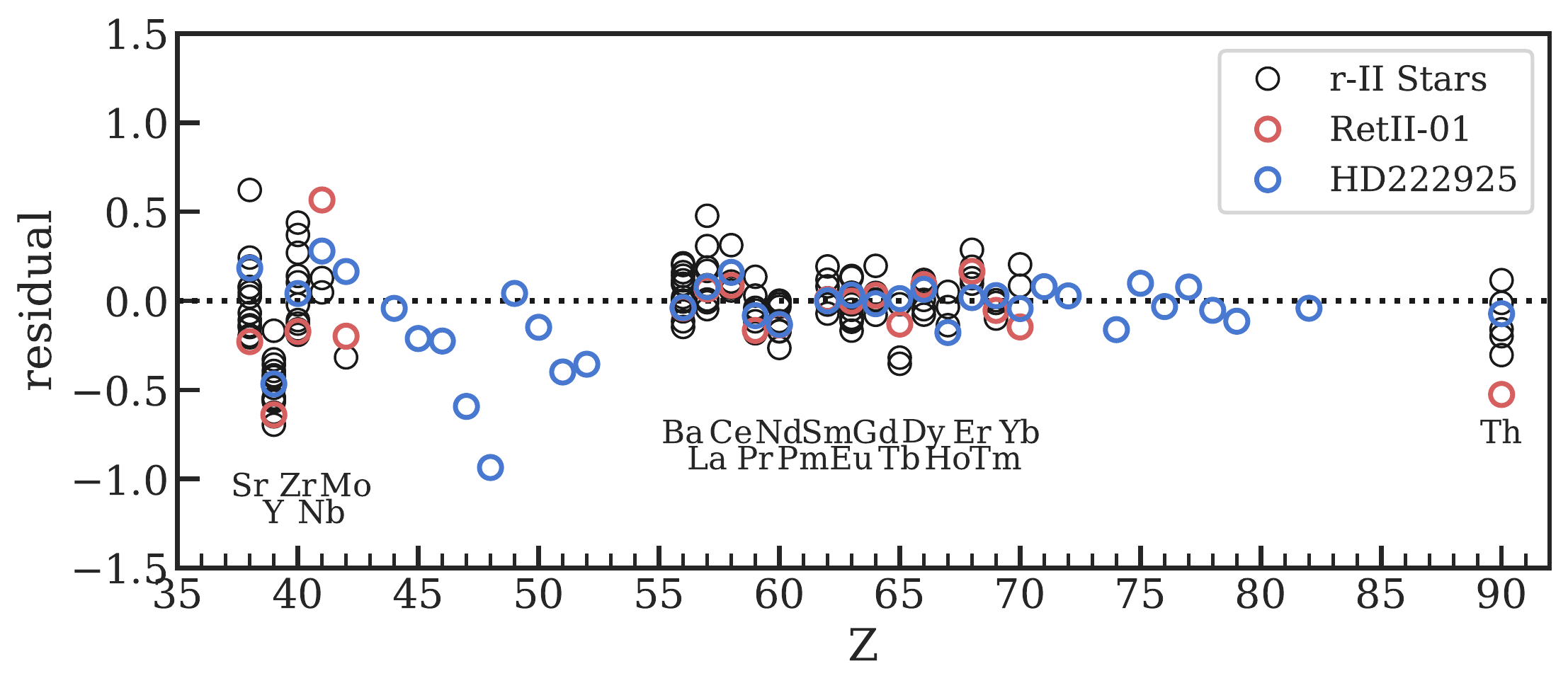}
    \caption{Residuals of stellar $r$-process element abundances subtracted from the solar $r$-process pattern \citep{Burris00}, for a variety of $r$-process enhanced ($r$-II) stars in the halo, the brightest star in the $r$-process dwarf galaxy Reticulum\,II \citep{Ji2018}, and the $r$-process  star HD\,222925 \citep{Roederer2022}.
    The residuals are calculated to minimize the absolute deviation for $56 \leq Z \leq 66$ (see \citealt{Ji2016c}).
    The universality of the $r$-process is reflected in the small amount of scatter of elements Ba to Yb, and when measurable in HD\,222925 it extends up to elements like Ir and Pt. On the contrary, the lighter neutron-capture elements Sr, Y, and Z show significant variations compared to the solar $r$-process pattern, with both higher and lower values.
    Variations in Th are also relatively large and representative of the actinide boost and deficiency phenomenon. Figure adapted and updated from \citet{Ji2016c}.}
    \label{fig:srresid}
\end{figure}

The behavior of the lighter neutron-capture elements, from strontium to barium, is not a universal pattern unlike what is found for the heavy $r$-process elements. 
Instead, significant variations, of over a factor of 10, are found in the ratio of these first $r$-process peak elements relative to the universal pattern occurring above barium in $r$-process enhanced stars \citep[e.g.,][]{Sneden2008,Ji2016c,Ji2018}.
This behavior is shown in Figure~\ref{fig:srresid}, which displays the stellar $r$-process residuals derived from subtracting off the solar $r$-process residuals \citep{Burris00} from the stellar $\log \epsilon$(X) abundances. It can clearly be seen that the scaled light neutron-capture element abundances (Sr, Y, Zr) vary from being significantly lower than the bulk of the scaled heavy-element pattern for some stars, to being much higher in other stars. In other words, there is substantial scatter in the observed [Sr/Eu] ratio, where Sr and Eu are representative elements for the light and heavy $r$-process elements, respectively. 
Note that it is important to keep in mind that the benchmark for comparisons, especially with theoretical predictions, has been the Sun's residual $r$-process pattern -- but the body of metal-poor stellar data does in fact not agree with the Sun among the lighter neutron-capture elements.

There are at least two possible explanations for the observed variations in the observed scatter among light neutron-capture element abundances.
First, the variation could result from the combination of two independent astrophysical sites, each producing a characteristic but stable abundance pattern in the first peak regime \citep[e.g.,][]{Travaglio04,Honda07,Montes07,SiqueiraMello2014,Shibagaki16}.
Alternatively, the variation could be due to intrinsic variability in $r$-process composition and yield within one astrophysical site, as is commonly expected in theoretical calculations  \citep[e.g.,][]{Wanajo14,Just15,Nishimura2015,Radice16,Wu2016}.
Mapping out these variations with more first peak elements in more $r$-process enhanced stars is ultimately needed to gain a more complete picture of light neutron-capture element production through the $r$-process.

An interesting future point of comparison could be multi-messenger astrophysics, where the ratio of light to heavy $r$-process elements can be observed directly through measuring detailed light curves of explosive transients \citep{kilpatrick17,Chornock2017,Tanaka2017,Tanvir2017}. Already, the one neutron star merger detected in gravitational waves, GW170817, whose electromagnetic counterpart was observed with telescopes has a composition inconsistent with the strongly enhanced $r$-II stars \citep{Ji2019b}.

\subsection{Actinide Element Variations and Nucleocosmochronometry}

The measured abundances of the actinide elements, thorium and uranium, are dominated by long-lived radioactive isotopes that decay over cosmic timescales. The species $^{238}$U and $^{232}$Th have half-lives of 4.5 and 14 billion years, respectively. Assuming metal-poor stars to be older than the Sun then implies that the actinide abundances would be lower than the Sun's values compared to a stable heavy element such as Eu. This behavior principally offers a way to age date old stars via this ``cosmo-chronometry", as long as initial production ratios (yields) of the progenitor $r$-process are known and reflective of the stars birth environment and conditions. This has so far proven very challenging despite site-independent yield calculations \citep{schatz02, hill17} that have been utilized to age date a number of metal-poor $r$-process stars \citep{Cayrel01,he1523,placco17}. Another challenge is related to observational uncertainties that quickly result in very large age uncertainties. Observational Th and U measurements are very difficult to achieve for a variety of reasons (e.g., requiring strong lines, a very high resolution, very high signal-to-noise spectrum, low stellar carbon abundance) but even reasonable uncertainties will result in several billion year age uncertainty showing once again how difficult cosmo-chronometry is observationally alone. Uncertainties in initial production ratios, easily a multiple of the observed uncertainties, add to this issue. See also \citet{Frebel2018} for further details.

To complicate matters further, about a quarter of metal-poor star with measurable Th (and U) abundances show a different behavior -- their actinide abundance are higher than those of the Sun's pattern \citep{holmbeck18}. This is illustrated in Figure~\ref{fig:srresid}. This phenomenon has been termed the ``actinide boost". This additional amount of actinides must come from a second and/or separate $r$-process or component that either occurred at a later time than the main/first event, or there must be significant variations in actinide production across $r$-processes and sites, perhaps due to fission cycling or other environmentally driven details \citep{holmbeck2019}.
It has also been found in a few individual stars in dwarf galaxies that there are  actinide deficient stars, with surprisingly low Th/Eu ratios \citep{Ji2018,Hansen2021}.

Generally, it would be of particular interest to measure actinide abundances in dwarf galaxy stars and star clusters, as one can obtain independent age measurements of these stellar systems from low-mass stellar evolution models. This would provide one way forward to calibrate $r$-process yield and site predictions in combination with an understanding of the gas mixing mass involved. So far, thorium has been measured in a number of stars in several dwarf galaxies: Reticulum~II \citep{Ji2018}, Ursa Minor \citep{aoki2007_cos82}, Fornax \citep{Reichert2021}, Sagittarius \citep{Hansen2020}, and the disrupted dwarf galaxy Indus \citep{Hansen2021}. However, uranium measurements will remain inaccessible due to the weakness of the strongest U line at 3859\,{\AA}.
The ages from cosmo-chronometry are consistent with the ages of the dwarf galaxies, but the large uncertainties from cosmo-chronometry make this comparison currently, unfortunately, fairly meaningless \citep{aoki2007_cos82}.

\begin{table}[]
    \scriptsize
    \setlength{\tabcolsep}{3pt}
    \centering
    \begin{tabular}{cc|r|rrr|rrr|rrr|rrr|rr}
Z & El & L09 & \multicolumn{3}{c|}{G99} & \multicolumn{3}{c|}{S08} & \multicolumn{3}{c|}{B14} & \multicolumn{3}{c|}{P20} & \multicolumn{2}{c}{R22} \\ \hline
&&$\log\epsilon$(X)$_{\odot}$&$s$&$r$&[X/Eu]&$s$&$r$&[X/Eu]&$s$&$r$&[X/Eu]&$s$&$r$&[X/Eu]&$r$&[X/Eu] \\ \hline
31 & Ga & $ 3.07$ & $ 2.96$ & $ 2.42$ & $-0.64$ & $ 2.83$ & $ 2.71$ & $-0.36$ & \nodata & \nodata & \nodata & $ 2.93$ & $ 2.51$ & $-0.54$ & $ 1.26$ & $-1.69$ \\
32 & Ge & $ 3.57$ & $ 3.41$ & $ 3.05$ & $-0.50$ & $ 3.29$ & $ 3.24$ & $-0.32$ & \nodata & \nodata & \nodata & $ 3.37$ & $ 3.13$ & $-0.42$ & $ 1.46$ & $-1.99$ \\
33 & As & $ 2.30$ & $ 1.88$ & $ 2.09$ & $-0.19$ & $ 1.63$ & $ 2.19$ & $-0.10$ & \nodata & \nodata & \nodata & $ 2.06$ & $ 1.92$ & $-0.36$ & $ 1.01$ & $-1.16$ \\
34 & Se & $ 3.34$ & $ 2.97$ & $ 3.09$ & $-0.23$ & $ 2.89$ & $ 3.14$ & $-0.19$ & \nodata & \nodata & \nodata & $ 2.91$ & $ 3.13$ & $-0.18$ & $ 2.62$ & $-0.59$ \\
35 & Br & $ 2.54$ & $ 1.77$ & $ 2.46$ & $-0.06$ & $ 1.53$ & $ 2.50$ & $-0.04$ & \nodata & \nodata & \nodata & $ 1.96$ & $ 2.41$ & $-0.11$ & \nodata & \nodata \\
36 & Kr & $ 3.26$ & $ 2.95$ & $ 2.96$ & $-0.28$ & $ 3.00$ & $ 2.91$ & $-0.34$ & $ 2.40$ & $ 3.19$ & $-0.04$ & $ 2.88$ & $ 3.02$ & $-0.21$ & \nodata & \nodata \\
37 & Rb & $ 2.37$ & $ 2.12$ & $ 2.01$ & $-0.34$ & $ 1.99$ & $ 2.13$ & $-0.23$ & $ 1.63$ & $ 2.28$ & $-0.06$ & $ 2.08$ & $ 2.06$ & $-0.29$ & \nodata & \nodata \\
38 & Sr & $ 2.87$ & $ 2.79$ & $ 2.12$ & $-0.73$ & $ 2.82$ & $ 1.91$ & $-0.95$ & $ 2.71$ & $ 2.36$ & $-0.48$ & $ 2.84$ & $ 1.78$ & $-1.07$ & $ 1.98$ & $-0.77$ \\
39 &  Y & $ 2.18$ & $ 2.06$ & $ 1.56$ & $-0.60$ & $ 2.03$ & $ 1.62$ & $-0.54$ & $ 2.03$ & $ 1.62$ & $-0.52$ & $ 2.07$ & $ 1.52$ & $-0.63$ & $ 1.04$ & $-1.01$ \\
40 & Zr & $ 2.54$ & $ 2.41$ & $ 1.97$ & $-0.56$ & $ 2.45$ & $ 1.82$ & $-0.71$ & $ 2.36$ & $ 2.07$ & $-0.45$ & $ 2.46$ & $ 1.80$ & $-0.72$ & $ 1.74$ & $-0.68$ \\
41 & Nb & $ 1.40$ & $ 1.34$ & $ 0.51$ & $-0.88$ & $ 1.23$ & $ 0.91$ & $-0.48$ & $ 1.15$ & $ 1.05$ & $-0.33$ & $ 1.22$ & $ 0.94$ & $-0.44$ & $ 0.71$ & $-0.57$ \\
42 & Mo & $ 1.80$ & $ 1.66$ & $ 1.23$ & $-0.56$ & $ 1.62$ & $ 1.32$ & $-0.47$ & $ 1.50$ & $ 1.49$ & $-0.28$ & $ 1.61$ & $ 1.34$ & $-0.44$ & $ 1.36$ & $-0.32$ \\
44 & Ru & $ 1.73$ & $ 1.26$ & $ 1.55$ & $-0.16$ & $ 1.28$ & $ 1.54$ & $-0.18$ & $ 1.22$ & $ 1.56$ & $-0.14$ & $ 1.30$ & $ 1.53$ & $-0.18$ & $ 1.32$ & $-0.28$ \\
45 & Rh & $ 1.08$ & $ 0.37$ & $ 0.98$ & $-0.08$ & $ 0.28$ & $ 1.00$ & $-0.07$ & $ 0.15$ & $ 1.02$ & $-0.03$ & $ 0.16$ & $ 1.02$ & $-0.03$ & $ 0.64$ & $-0.31$ \\
46 & Pd & $ 1.64$ & $ 1.31$ & $ 1.36$ & $-0.26$ & $ 1.28$ & $ 1.39$ & $-0.24$ & $ 1.20$ & $ 1.44$ & $-0.17$ & $ 1.30$ & $ 1.37$ & $-0.24$ & $ 1.05$ & $-0.46$ \\
47 & Ag & $ 1.20$ & $ 0.54$ & $ 1.09$ & $-0.09$ & $ 0.53$ & $ 1.10$ & $-0.09$ & $ 0.24$ & $ 1.15$ & $-0.02$ & $ 0.52$ & $ 1.10$ & $-0.08$ & $ 0.44$ & $-0.64$ \\
48 & Cd & $ 1.70$ & $ 1.42$ & $ 1.37$ & $-0.31$ & $ 1.41$ & $ 1.38$ & $-0.30$ & $ 1.37$ & $ 1.42$ & $-0.25$ & $ 1.45$ & $ 1.33$ & $-0.34$ & $ 0.34$ & $-1.23$ \\
49 & In & $ 0.74$ & $ 0.28$ & $ 0.56$ & $-0.17$ & $ 0.25$ & $ 0.57$ & $-0.16$ & $ 0.21$ & $ 0.59$ & $-0.13$ & $ 0.33$ & $ 0.53$ & $-0.19$ & $ 0.51$ & $-0.11$ \\
50 & Sn & $ 2.06$ & $ 1.89$ & $ 1.57$ & $-0.47$ & $ 1.94$ & $ 1.42$ & $-0.63$ & \nodata & \nodata & \nodata & $ 1.91$ & $ 1.52$ & $-0.52$ & $ 1.39$ & $-0.55$ \\
51 & Sb & $ 1.01$ & $ 0.58$ & $ 0.80$ & $-0.19$ & $ 0.23$ & $ 0.93$ & $-0.07$ & $ 0.30$ & $ 0.91$ & $-0.07$ & $ 0.40$ & $ 0.88$ & $-0.10$ & $ 0.37$ & $-0.51$ \\
52 & Te & $ 2.18$ & $ 1.46$ & $ 2.09$ & $-0.07$ & $ 1.46$ & $ 2.09$ & $-0.08$ & \nodata & \nodata & \nodata & $ 1.49$ & $ 2.08$ & $-0.08$ & $ 1.63$ & $-0.43$ \\
53 &  I & $ 1.55$ & $ 0.91$ & $ 1.44$ & $-0.10$ & $ 0.29$ & $ 1.53$ & $-0.02$ & $ 0.13$ & $ 1.53$ & $ 0.01$ & $ 0.06$ & $ 1.54$ & $ 0.01$ & \nodata & \nodata \\
54 & Xe & $ 2.25$ & $ 1.85$ & $ 2.02$ & $-0.20$ & $ 1.55$ & $ 2.15$ & $-0.09$ & \nodata & \nodata & \nodata & $ 1.51$ & $ 2.16$ & $-0.07$ & \nodata & \nodata \\
55 & Cs & $ 1.08$ & $ 0.30$ & $ 1.00$ & $-0.06$ & $ 0.26$ & $ 1.01$ & $-0.06$ & $ 0.21$ & $ 1.02$ & $-0.04$ & $ 0.28$ & $ 1.01$ & $-0.05$ & \nodata & \nodata \\
56 & Ba & $ 2.16$ & $ 2.09$ & $ 1.31$ & $-0.83$ & $ 2.09$ & $ 1.32$ & $-0.83$ & $ 2.09$ & $ 1.32$ & $-0.81$ & $ 2.11$ & $ 1.23$ & $-0.90$ & $ 1.26$ & $-0.78$ \\
57 & La & $ 1.17$ & $ 0.99$ & $ 0.71$ & $-0.45$ & $ 1.05$ & $ 0.56$ & $-0.60$ & $ 1.05$ & $ 0.56$ & $-0.58$ & $ 1.07$ & $ 0.47$ & $-0.68$ & $ 0.51$ & $-0.54$ \\
58 & Ce & $ 1.58$ & $ 1.49$ & $ 0.86$ & $-0.70$ & $ 1.49$ & $ 0.86$ & $-0.71$ & $ 1.50$ & $ 0.79$ & $-0.77$ & $ 1.51$ & $ 0.77$ & $-0.79$ & $ 0.85$ & $-0.61$ \\
59 & Pr & $ 0.75$ & $ 0.30$ & $ 0.55$ & $-0.18$ & $ 0.44$ & $ 0.45$ & $-0.28$ & $ 0.44$ & $ 0.45$ & $-0.27$ & $ 0.47$ & $ 0.41$ & $-0.31$ & $ 0.22$ & $-0.40$ \\
60 & Nd & $ 1.44$ & $ 1.20$ & $ 1.08$ & $-0.35$ & $ 1.20$ & $ 1.07$ & $-0.36$ & \nodata & \nodata & \nodata & $ 1.22$ & $ 1.04$ & $-0.38$ & $ 0.88$ & $-0.44$ \\
62 & Sm & $ 0.92$ & $ 0.39$ & $ 0.77$ & $-0.14$ & $ 0.40$ & $ 0.77$ & $-0.14$ & $ 0.44$ & $ 0.75$ & $-0.14$ & $ 0.44$ & $ 0.75$ & $-0.15$ & $ 0.62$ & $-0.18$ \\
63 & Eu & $ 0.50$ & $-0.91$ & $ 0.49$ & $ 0.00$ & $-1.18$ & $ 0.49$ & $ 0.00$ & $-0.72$ & $ 0.48$ & $ 0.00$ & $-0.81$ & $ 0.48$ & $ 0.00$ & $ 0.38$ & $ 0.00$ \\
64 & Gd & $ 1.07$ & $ 0.37$ & $ 0.97$ & $-0.08$ & $ 0.31$ & $ 0.98$ & $-0.07$ & $ 0.25$ & $ 0.99$ & $-0.04$ & $ 0.29$ & $ 0.98$ & $-0.06$ & $ 0.82$ & $-0.12$ \\
65 & Tb & $ 0.31$ & $-0.97$ & $ 0.29$ & $-0.01$ & $-0.89$ & $ 0.28$ & $-0.02$ & $-0.78$ & $ 0.28$ & $-0.01$ & $-0.83$ & $ 0.28$ & $-0.01$ & $ 0.18$ & $-0.01$ \\
66 & Dy & $ 1.12$ & $ 0.10$ & $ 1.07$ & $-0.03$ & $ 0.19$ & $ 1.06$ & $-0.05$ & $ 0.29$ & $ 1.05$ & $-0.04$ & $ 0.30$ & $ 1.04$ & $-0.05$ & $ 1.01$ & $ 0.02$ \\
67 & Ho & $ 0.47$ & $-0.64$ & $ 0.43$ & $-0.02$ & $-0.70$ & $ 0.44$ & $-0.02$ & $-0.61$ & $ 0.43$ & $-0.01$ & $-0.66$ & $ 0.44$ & $-0.01$ & $ 0.12$ & $-0.23$ \\
68 & Er & $ 0.92$ & $ 0.12$ & $ 0.85$ & $-0.06$ & $ 0.11$ & $ 0.85$ & $-0.06$ & $ 0.19$ & $ 0.83$ & $-0.06$ & $ 0.19$ & $ 0.83$ & $-0.07$ & $ 0.73$ & $-0.07$ \\
69 & Tm & $ 0.12$ & $-0.67$ & $ 0.04$ & $-0.06$ & $-0.67$ & $ 0.04$ & $-0.07$ & $-0.92$ & $ 0.08$ & $-0.01$ & $-0.77$ & $ 0.06$ & $-0.04$ & $-0.09$ & $-0.09$ \\
70 & Yb & $ 0.92$ & $ 0.46$ & $ 0.73$ & $-0.17$ & $ 0.42$ & $ 0.75$ & $-0.16$ & $ 0.53$ & $ 0.69$ & $-0.20$ & $ 0.55$ & $ 0.68$ & $-0.22$ & $ 0.55$ & $-0.24$ \\
71 & Lu & $ 0.09$ & $-0.61$ & $-0.01$ & $-0.08$ & $-0.64$ & $ 0.00$ & $-0.08$ & $-0.59$ & $-0.01$ & $-0.07$ & $-0.61$ & $-0.01$ & $-0.08$ & $-0.04$ & $-0.01$ \\
72 & Hf & $ 0.71$ & $ 0.45$ & $ 0.36$ & $-0.33$ & $ 0.39$ & $ 0.43$ & $-0.27$ & $ 0.49$ & $ 0.31$ & $-0.38$ & $ 0.48$ & $ 0.32$ & $-0.37$ & $ 0.32$ & $-0.27$ \\
73 & Ta & $-0.17$ & $-0.47$ & $-0.46$ & $-0.28$ & $-0.56$ & $-0.40$ & $-0.22$ & $-0.50$ & $-0.44$ & $-0.25$ & $-0.47$ & $-0.47$ & $-0.28$ & \nodata & \nodata \\
74 &  W & $ 0.65$ & $ 0.42$ & $ 0.25$ & $-0.38$ & $ 0.37$ & $ 0.32$ & $-0.32$ & $ 0.44$ & $ 0.23$ & $-0.39$ & $ 0.43$ & $ 0.25$ & $-0.38$ & $ 0.02$ & $-0.50$ \\
75 & Re & $ 0.27$ & $-0.44$ & $ 0.18$ & $-0.08$ & $-0.74$ & $ 0.23$ & $-0.04$ & $-0.53$ & $ 0.20$ & $-0.05$ & $-0.54$ & $ 0.20$ & $-0.05$ & $ 0.16$ & $ 0.01$ \\
76 & Os & $ 1.34$ & $ 0.40$ & $ 1.29$ & $-0.04$ & $ 0.28$ & $ 1.30$ & $-0.03$ & $ 0.42$ & $ 1.29$ & $-0.03$ & $ 0.35$ & $ 1.29$ & $-0.03$ & $ 1.17$ & $-0.05$ \\
77 & Ir & $ 1.34$ & $ 0.24$ & $ 1.30$ & $-0.02$ & $-0.59$ & $ 1.33$ & $ 0.00$ & $-0.46$ & $ 1.33$ & $ 0.02$ & $-0.61$ & $ 1.33$ & $ 0.02$ & $ 1.28$ & $ 0.07$ \\
78 & Pt & $ 1.62$ & $-0.01$ & $ 1.60$ & $ 0.01$ & $ 0.33$ & $ 1.59$ & $-0.01$ & $ 0.40$ & $ 1.59$ & $-0.00$ & $ 0.51$ & $ 1.58$ & $-0.01$ & $ 1.45$ & $-0.04$ \\
79 & Au & $ 0.80$ & $-0.88$ & $ 0.79$ & $ 0.01$ & $-0.47$ & $ 0.78$ & $-0.01$ & $-0.41$ & $ 0.77$ & $-0.00$ & $-0.44$ & $ 0.77$ & $-0.00$ & $ 0.53$ & $-0.15$ \\
80 & Hg & $ 1.17$ & $ 0.98$ & $ 0.72$ & $-0.43$ & $ 0.93$ & $ 0.80$ & $-0.36$ & $ 0.88$ & $ 0.86$ & $-0.28$ & $ 0.92$ & $ 0.82$ & $-0.33$ & \nodata & \nodata \\
81 & Tl & $ 0.77$ & $ 0.62$ & $ 0.23$ & $-0.52$ & $ 0.58$ & $ 0.32$ & $-0.44$ & $ 0.62$ & $ 0.24$ & $-0.51$ & $ 0.65$ & $ 0.15$ & $-0.60$ & \nodata & \nodata \\
82 & Pb & $ 2.03$ & $ 1.98$ & $ 1.04$ & $-0.97$ & $ 1.93$ & $ 1.36$ & $-0.66$ & $ 1.97$ & $ 1.14$ & $-0.87$ & $ 1.95$ & $ 1.25$ & $-0.76$ & $ 1.14$ & $-0.77$ \\
83 & Bi & $ 0.65$ & $ 0.45$ & $ 0.21$ & $-0.42$ & $ 0.20$ & $ 0.46$ & $-0.18$ & $-0.04$ & $ 0.55$ & $-0.07$ & $-0.02$ & $ 0.54$ & $-0.08$ & \nodata & \nodata \\
90 & Th & $ 0.15$ & \nodata & $ 0.15$ & $ 0.02$ & \nodata & $ 0.15$ & $ 0.01$ & \nodata & \nodata & \nodata & \nodata & $ 0.15$ & $ 0.02$ & $-0.06$ & $-0.09$ \\
92 &  U & $-0.11$ & \nodata & $-0.11$ & $ 0.02$ & \nodata & $-0.11$ & $ 0.01$ & \nodata & \nodata & \nodata & \nodata & $-0.11$ & $ 0.02$ & \nodata & \nodata \\
    \end{tabular}
    \caption{Solar patterns for $s$-process and $r$-process from \citet{Goriely1999} (G99), \citet{Sneden2008} (S08), \citet{Bisterzo2014} (B14), and \citet{Prantzos2020} (P20); along with the $r$-process pattern for HD\,222925 from \citet{Roederer2022} (R22).
    The solar abundance is adopted from \citet{lodders09} meteoritic abundances (L09) but renormalized to $\log\epsilon$(X) units using the \citet{Asplund2009} solar silicon abundance of $\log\epsilon$(Si)$_\odot=7.51$.
    The offset is because meteoritic abundances are normalized to N(Si)$=6$ while the spectroscopist's $\log\epsilon$(X) values are normalized to N(H)$=12$. The $s$ and $r$ columns are also given in $\log\epsilon$(X) units, while the [X/Eu] column indicates the abundance of each element relative to solar in standard spectroscopist's notation.
    Note that even [Ba/Eu]$_r$ varies from $-0.78$ to $-0.90$ across these five different $r$-process patterns despite using the same overall solar isotopic abundance, which emphasizes the model-dependence of the solar $r$-process residual.
    }
    \label{tab:solarrproc}
\end{table}

\section{Candidate Astrophysical Sites of the $r$-process}\label{candidates}

The fundamental question at hand is about the origin of heavy elements synthesized in the rapid neutron-capture elements: What are possible astrophysical sites, and what is their relative importance over cosmic history?
A number of sites have been proposed theoretically to explain the operation of the $r$-process.
Since  $r$-process nucleosynthesis requires a very high density of free neutrons, it can essentially only happen in events associated with the birth and death of compact neutron stars.

Conditions for an $r$-process to operate must include a high neutron-to-seed ratio. The most important quantity for describing the neutron-to-seed ratio is the electron fraction $Y_e = n_e/n$. Since the ejecta is net neutral, $Y_e$ is  equal to the fraction of protons, and $1-Y_e$ is the fraction of neutrons. Thus, a lower $Y_e$ means that the ejecta will be more neutron-rich.
A broad rule of thumb about $r$-process nucleosynthesis calculations is that an electron fraction of $Y_e > 0.25$ would result in the build up of elements of the first peak region, $0.10 < Y_e < 0.25$ would lead to elements partaking in the universal $r$-process pattern between the second and third $r$-process peak as well as some actinide elements, and $Y_e < 0.10$ would be especially responsible for the production of the actinide elements \citep[e.g.,][]{Lippuner15,Wu2016,holmbeck2019}.
However, there are many exceptions to this general rule due to the exact thermodynamic properties of the ejecta \citep[e.g.,][]{Farouqi10,Wanajo14,Mumpower2016,Curtis2021}.
Note that when discussing the origin of $r$-process elements, most people implicitly assume the question concerns solely the origin of $r$-process elements beyond the second $r$-process peak: The rare earth elements, the third $r$-process peak, and the actinides. Focus on these heavier elements stems from the fact that elements in the first $r$-process peak can also be produced in a much larger variety of ways, such as the weak $s$-process or $\nu p$ process \citep{Sukhbold2016, Frohlich2006}.

Here, three broad categories of the most commonly invoked astrophysical sites for the $r$-process to occur are considered \citep{Cowan2021}:
Common core-collapse supernovae (more than 1-10\%), e.g., in neutrino-driven winds \citep[e.g.,][]{Woosley92,woosley94,Farouqi10,Arcones11} or during electron-capture supernovae \citep{Ning2007,Janka2008,Wanajo2011};
rare core-collapse supernovae (less than 1\% occurrence rate), such as in magnetorotationally driven jets \citep{Winteler2012,Nishimura2015,Nishimura2017,Mosta2018} or in collapsar disk winds \citep{Pruet2004,Surman2011,Siegel2019,Miller2020};
and binary neutron star mergers, including both neutron star-neutron star pairs and black hole-neutron star mergers \citep[e.g.,][]{Lattimer74,Lattimer76,Lattimer77,Surman2008,Hotokezaka2013,Goriely11,Goriely2013,rosswog14,Wanajo14,Eichler2015,Just15,Wu2016,Lippuner2017,Curtis2021}.

Below is a brief summary of the results of theoretical studies on four major questions:
How often does it occur -- what is the event rate?
How quickly does it happen after star formation -- what is the delay time distribution?
How much $r$-process material is made -- What is the event yield? and
What is the distribution of elements made -- What is the $r$-process composition?

\subsection{Common core-collapse supernovae}

Since \citet{Burbidge57}, core-collapse supernovae have been regarded as a promising site for $r$-process nucleosynthesis.
But it has also been clear since \citet{mcwilliam98} that supernovae could not all produce the same amount of $r$-process elements. Yet a specific range of stellar masses, upon their explosion, could principally be able to produce all the $r$-process elements through the second and third peak. This would imply the operation of $r$-process nucleosynthesis in a relatively large fraction (1-10\%) of core-collapse supernovae \citep[e.g.,][]{wanajo_ishimaru}.

Hypothetical mechanisms for producing the full slew of the $r$-process elements include neutrino-driven winds emanating from the proto-neutron star \citep[e.g.,][]{takahashi94,Woosley92,woosley94,Qian96,Farouqi10} and electron-capture supernovae of $8-10 M_\odot$ stars \citep[e.g.,][]{Ning2007,Janka2008,wanajo_rs2006}.
Due to the short lifetimes of their progenitor stars, these common core-collapse supernovae would produce a small amount of $r$-process elements essentially immediately after an episode of star formation from which they originated.
However, modern calculations show that these mechanisms tend not to produce a full $r$-process, though they can produce a limited (weak) $r$-process up to the first $r$-process peak \citep[e.g.,][]{Arcones07,Arcones11,Wanajo2011}.
Thus, theoretically such sites are disfavored as the source of the heaviest $r$-process elements.

Note that there is evidence that such a limited $r$-process (also often called the light element primary process, \citealt{Travaglio04,Montes07,Hansen12}) does exist in nature. A group of metal-poor stars, now termed limited $r$-process stars (e.g., \citealt{Frebel2018, Hansen2018}), qualitatively shows this type of abundance drop-off with increasing atomic number, as expected for a limited (weak) $r$-process \citep{Honda04,Honda07}. Following the abundance definition for these limited-$r$ stars, they need to fulfill $\mbox{[Eu/Fe]} < 0.3$, $\mbox{[Sr/Ba]} > 0.5$, and  $\mbox{[Sr/Eu]} > 0.0$.

\subsection{Rare core-collapse supernovae}

Given the difficulties of the common core-collapse supernovae to produce a full $r$-process, contemporary models of $r$-process nucleosynthesis in core-collapse supernovae invoke extreme physical conditions that may occur rarely, in less than 1\% of core-collapse supernova explosions (if physical indeed).
Currently, the most popular models are collapsar disk winds \citep{Pruet2004,Surman2011,Siegel2019,Miller2020} and magnetorotationally driven jet supernovae \citep{Winteler2012,Nishimura2015,Mosta2018,Halevi2018}, which both involve high magnetic fields and very rapid rotation. Note these two mechanisms are not mutually exclusive, and in principle both could happen in the same supernova.
Like the common core-collapse supernovae, it is currently rather uncertain whether these rare core-collapse supernovae are even capable of synthesizing the heaviest $r$-process elements, due to the complex physics involved (e.g. neutrino transport, radiation magnetohydrodynamics, general relativity).
However, if suitable conditions for $r$-process production can be achieved, the $r$-process yields can be relatively high ($0.01-1 M_\odot$) and they would also occur promptly after star formation due to the short lives of the progenitors.

\subsection{Neutron star mergers}

Neutron star mergers  result from the inspiraling of two compact objects that have been orbiting each other and losing energy through gravitational waves. The two subclasses are binary neutron star mergers (BNS) and black hole-neutron star mergers (BHNS).
The gravitational wave event GW170817 and its associated electromagnetic kilonova counterpart AT2017gfo \citep{LIGOGW170817a,LIGOGW170817b} have provided conclusive evidence that binary neutron star mergers can produce heavy $r$-process elements \citep{Arcavi2017,Chornock2017,Coulter2017,Cowperthwaite2017,drout17,Evans2017,Kasliwal2017,kilpatrick17,McCully2017,Nicholl2017,Pian2017,shappee17,Smartt2017,Tanvir2017,Troja2017}. This event remains the only direct observation of $r$-process nucleosynthesis in the cosmos, though only one individual element (strontium) was identified \citep{Watson2019}.
These neutron star mergers produce a moderately high amount of $r$-process ($\sim 0.01 M_\odot$) material, and it is expected that they will produce a variable ratio of light  and heavy $r$-process elements \citep[e.g.,][]{Wu2016,Lippuner2017,Ji2019b}.
Due to the time needed to inspiral due to a continued loss of energy through gravitational radiation, the actual merger events are expected to be delayed relative to star formation episode that created their massive progenitor stars \citep[e.g.,][]{Dominik12}.
Additionally, binary neutron stars can experience a substantial velocity kick of a few hundred km/s during the formation of each of the neutron stars \citep[e.g.,][]{Willems2004,Dominik12,Fong2013,Beniamini2016,Bramante2016}.

While neutron star mergers are the only definite source of $r$-process elements, it is still not clear how important they might be for the buildup of $r$-process elements in the Galaxy, and in particular, for understanding the composition of stars like our Sun. Reasons include that the delayed $r$-process production may well occur only after most stars in a galaxy have already formed, and/or that the enrichment in $r$-process elements may occur far away from the original star-forming regions if the system experienced strong velocity kicks.

One other important caveat is that there is currently significant debate about how quickly neutron stars can actually merge. Most expectations are that a typical merging time is distributed as $t_{\rm merge} \propto t^{-1}$ with a minimum delay time of 10-100\,Myr \citep[e.g.,][]{Dominik12,Belczynski2018,Chruslinska2018,Neijssel2019}, but several authors have argued that a substantial fraction of neutron star binaries should merge much faster, on the order of 1\,Myr after both supernovae in the binary system have exploded \citep[e.g.,][]{Beniamini2019,Safarzadeh2019}.

\subsection{Distinguishing factors}

Comparing these three classes of sites, common core-collapse supernovae are theoretically disfavored due to their inability to synthesize the heavy $r$-process elements (as well as observationally ruled out by dwarf galaxies; Section~\ref{sec:ufd}).
The other two classes, rare core-collapse supernovae and neutron star mergers, have very similar event rates and $r$-process yields. These rare events have very few observational constraints, and if indeed physically occurring in the universe, they operate under extreme multi-physics circumstances. Given significant theoretical uncertainties in the predicted electron fraction distributions, the uncertainties in nuclear experimental data along the $r$-process path, and few direct observational constraints, both of the rare $r$-process production sites should currently be considered as able to reproduce the typical $r$-process pattern, certainly for elements barium and above. Additional differences may be present on an isotopic level but that unfortunately cannot be constrained with observations of metal-poor stars.

Thus, the primary available distinguishing factor between rare core-collapse supernovae and neutron star mergers as the primary source of $r$-process elements in the universe is the \emph{delay time distribution}, i.e. how long it takes after a burst of star formation in a given system for $r$-process elements to be synthesized.
Delay times for core-collapse supernovae are simply the lifetimes of massive stars, which are extremely short (or ``prompt''), $10-20$\,Myr.
Delay times for neutron star mergers are typically extended (or ``delayed''), ranging from ${\sim}30$\,Myr to several Gyr \citep[e.g.,][]{Belczynski2018,Neijssel2019}.
The existence of $r$-process enhanced metal-poor stars thus traditionally led astronomers to prefer a prompt enrichment channel, such as core-collapse supernovae as the origin of $r$-process elements \citep[e.g.,][]{Qian00,Argast04,wanajo_ishimaru}. 

However, it is important to recall that inferring the delay time distribution is not independent of the astrophysical context in which an $r$-process event occurs.
As the most obvious example, the correspondence between metallicity and time (often called the age-metallicity relation) varies substantially depending on the mass of a galaxy that a star forms in \citep[e.g.,][]{Ishimaru15,Ji2016b}. Stars forming in low mass galaxies take much longer to reach a level of even [Fe/H] $= -3$ compared to stars forming in high mass galaxies.
Furthermore, the detailed dynamics of how metals from an explosive event mix into the surrounding interstellar gas (``metal mixing'') can introduce inhomogeneities, such that coeval stars have in fact a range of metallicities, or conversely, that a single metallicity corresponds to a range of ages \citep[e.g.,][]{Webster14,Emerick2019,Ji2022}.
Additionally, the $r$-process elements are synthesized at high energy, and it takes time to trap the material and cool it back down to star-forming gas in a galaxy's interstellar medium \citep{Shen2015,Hirai15,Naiman2018,Schonrich2019,Cote2019,Naidu2022,Amend2022,vandeVoort2020,vandeVoort2022}
Thus, $r$-process enrichment should vary across time and also between galaxies with properties and environments that include different star formation histories and masses.
Unfortunately, the study of metal-poor Galactic halo stars limits the exploration of these important issues; they cannot provide any information of environment of host galaxy properties because these halo stars originate from a variety of dwarf galaxies of different masses that were accreted by the Milky Way. Other stars are needed to probe both $r$-process abundance signatures as well as reveal environmental information about their birth conditions.

\section{Dwarf Galaxies: Nature's Best $R$-process Laboratories}\label{sec:dwarfgals}

The Milky Way's dwarf satellite galaxies have been extensively studied over the past half century (see e.g. \citealt{Tolstoy2009} for a comprehensive summary, and also \citealt{Simon2019} for more recent results). Given the specific focus here on $r$-process observations in dwarf galaxies, here is a brief summary of the state of the art, including providing information on the basic properties of dwarf galaxies as understanding dwarf galaxy environments has become increasingly important for learning about the site and operation of the $r$-process.

There are currently ${\sim60}$ dwarf galaxy satellites of the Milky Way \citep{Simon2019}.
Their basic properties are tabulated in Table~\ref{tab:dwarfprop} and show some important quantities in Figure~\ref{fig:dwarfprop}.
Only two of these galaxies contain gas and are star-forming (the Large and Small Magellanic Clouds, LMC and SMC). All of the other galaxies are so-called dwarf spheroidal (dSph) galaxies, which contain no neutral gas and look morphologically like a diffuse cloud of stars.
It is thus useful here to define ``galaxy,'' as opposed to other gravitationally bound stellar systems such as star clusters \citep{Willman2012}. A galaxy forms at the center of a dark matter halo, so its gravitational mass is 10-100 times higher than its stellar mass. Even galaxies with only a few thousand solar masses of stars have their own extensive dark halo and thus represent the lower mass end of galaxy formation. On the contrary, a star cluster is not embedded in a dark halo, even the most massive ones containing several million solar masses of stars.
Twelve of the Milky Way satellites are still candidate galaxies at the time of writing, meaning that there is not yet enough data to conclusively distinguish them from star clusters.

Some confusing naming conventions should also be clarified. Dwarf galaxies are typically named after the International Astronomical Union-defined constellation where they were originally discovered, followed by a roman numeral if multiple galaxies or candidate galaxies or structures are in the same constellation on the sky. For example,  Reticulum\,II is the 2nd galaxy-like structure found within the Reticulum constellation. The first galaxy discovered typically has the roman numeral dropped (e.g., Carina\,I = Carina) as is especially the case with the longer known systems. However, upon initial discovery, it can be unclear if a system is ultimately confirmed to be a galaxy or rather a star cluster, as the structure needs to be followed up with spectroscopy to confirm a high mass-to-light ratio and metallicity dispersion. As a result, galaxy names do not always follow this rule. For example, Segue 1 and 2 were named after the SEGUE survey, instead of a constellation, and were only later confirmed to actually be galaxies. Reticulum\,I has been shown to be a globular cluster, but Reticulum~II is a dwarf galaxy. Once names are given, they are typically kept.

\subsection{The Need to Study Dwarf Galaxies to Learn about the $R$-process}

Contrary to the availability of relatively bright and close-by halo stars with observable $r$-process enhancements, stars in dwarf galaxies are much, much fainter. This is mostly due to the large distances of often up to 250\,kpc of these system, leaving only the brightest red giant stars available for detailed spectroscopic observations. These brightest stars  are of 16th to 19th magnitude, where 19th magnitude is at the edge of what is technically possible to observe with high-resolution spectroscopy even with the largest telescopes.
Every single star thus takes at least several hours if not an entire night or more to observe. 

However, stars in dwarf galaxies provide a powerful advantage over Galactic halo stars:
A dwarf galaxy provide a collection of stars with a known common and correlated formation history. This includes environmental information related to metal mixing, gas dilution mass, and star formation history. The observed stars are still within their birth environment, which means that the chemical enrichment that took place prior to their birth can be reconstructed more quantitatively. Knowing about the environment helps to break degeneracies of galaxy evolution. At fixed [X/H], the number of enriching sources is completely degenerate with the gas mass of the system. But within the same galaxy, the gas mass is constant, thus offering a way to constrain the enrichment history. 
This is essential when trying to understand the signature and origin scenarios of different nucleosynthesis processes. Since the Galactic halo formed from a combination of dissolved dwarf galaxies as part of the hierarchical assembly of the Milky Way, any given halo star likely did not originate from within the Galaxy but in a small dwarf galaxy long ago. Hence, any information on the birth gas cloud conditions was lost upon the accretion of the erstwhile host galaxy.

While element nucleosynthesis is entirely governed by nuclear physics processes, the initial conditions for it to occur are determined by the astrophysical site and associated environmental situation at the time of the event. On one hand, this complicates the interpretation of the observed signatures, but on the other hand, it also provides an opportunity to obtain information that might provide insight into the astrophysical site and conditions. 

Regarding $r$-process nucleosynthesis, stars located in dwarf galaxies are ideally suited for studying individual, clean nucleosynthesis processes, since they still reside in the environment in which the $r$-process event occurred. Besides the yield constraints, the dwarf galaxy environment specifically provides means to constrain different production sites in terms of their astrophysics delay times and and type of yields. With chemical evolution models, varying delay times can be taken into account when  predicting observed trends and abundance ratios pertaining to contributions by prompt and delayed sources. However, the sites associated with those timescales remain indistinguishable at present. More data, likely to come in the next decades, will hopefully shed more light on these questions.

\subsection{$R$-process Observables in Dwarf Galaxies}

Of all the neutron-capture elements, only three are usually available to measure in dwarf galaxies: strontium (Sr), barium (Ba), and europium (Eu), which trace the first $r$-process peak, the second $r$-process peak, and the rare earth elements, respectively.
This is because in low-metallicity stars ($\mbox{[Fe/H]} < -2$), the absorption lines of other neutron-capture elements are intrinsically much weaker and as such usually undetectable in spectrum of low or moderate quality.
At metallicities of $\mbox{[Fe/H]} \gtrsim -2.0$, yttrium (Y) becomes measurable in place of Sr, and lanthanum (La) in place of Ba. With increasing [Fe/H] many other elements also become  measurable (especially Zr, Ce, Nd, Dy), especially if their abundances are enhanced.
Indeed, in a few rare cases of highly $r$-process enhanced stars, it has been possible to measure the actinide element thorium (Th) \citep{aoki2007_cos82,Ji2018,Hansen2018,Hansen2021,Reichert2021}.

Still, measuring neutron-capture element abundances in dwarf galaxies is hard.
Since dwarf galaxies are at least 20 kiloparsecs away, it is currently only feasible to study red giant stars, and unfortunately most neutron-capture element absorption lines occur at relatively low (blue) wavelengths where these stars produce little flux and spectrograph detectors have poor quantum efficiency. Additionally, moderately high spectral resolution ($R = \lambda/\Delta \lambda \gtrsim 25,000$) is required to detect any of the intrinsically weak $r$-process element absorption lines.
This generally limits observations to the available brighter stars with $V < 19$, and requires, despite the above challenges, reaching down to wavelengths of 4000\,{\AA} to measure at least Sr (4077\,{\AA} and 4215\,{\AA}) and Eu (4129\,{\AA} and other lines further to the blue and red).
The easiest of the three element to measure is Ba because has multiple strong lines at relatively red wavelengths (4554-6497\,{\AA}). Recently, it has been shown that Ba can be accurately measured at lower spectral resolutions of $R \sim 6000$ by simultaneously fitting all the Ba lines \citep{Duggan2018}. Thus, in many dwarf galaxy studies, Ba is the only neutron-capture element available.

In general, dwarf galaxies are enriched by a combination of the $r$-process and $s$-process.
Before studying $r$-process nucleosynthesis and its history, the $s$-process component must be removed. This is typically accomplished by measuring both Ba and Eu abundances, then assuming that the solar $r$-process pattern is universal from the second to the third peak ([Ba/Eu] $= -0.8$, \citealt{Sneden2008}), and that any excess Ba is due to the $s$-process \citep[e.g.,][]{Duggan2018,Skuladottir2019,delosReyes2022}.
In principle, this fundamentally limits our ability to use dwarf galaxies to study variations in the $r$-process composition.
However, the most metal-poor stars in the least massive dwarf galaxies can be assumed to have formed early enough that $s$-process contamination of the system is unlikely. Measurements made for some of these stars have confirmed that deviations from the universal $r$-process pattern are at the same level as measurement uncertainties. Thus, when considering average abundance trends, it is a reasonable procedure to do this subtraction, though it precludes studying variations in the detailed $r$-process composition.

Overall, it is thus possible to study the (more) detailed $r$-process patterns of dwarf galaxy stars only in the rare cases where the stars are unusually close, and only with a very large investment of telescope time \citep[e.g.,][]{Ji2018,Hansen2021}.
However bulk distributions of a few element ratios are currently achievable, such as [Ba/Eu] and [Sr/Eu].

\subsection{General Properties of Dwarf Galaxies}

\begin{table}[]
    \centering
    \scriptsize
    \begin{tabular}{l|llrrrl}
Galaxy Name & R.A. & Decl. & $M_V$ & [Fe/H] & $t_{\rm SF}$ & References \\ \hline
Large~Magellanic~Cloud & 05:23:34.6 & $-$69:45:22 & $-18.12$ & $-0.50$ & $ 13.8$ & 33,33,33,NA \\
Small~Magellanic~Cloud & 00:52:44.8 & $-$72:49:43 & $-16.83$ & $-1.00$ & $ 13.8$ & 33,33,33,NA \\
Sagittarius            & 18:54:59.2 & $-$30:27:38 & $-13.50$ & $-0.53$ & $ 10.4$ & 31,31,34,47 \\
Fornax                 & 02:39:50.0 & $-$34:29:59 & $-13.34$ & $-1.07$ & $ 11.6$ & 35,35,21,47 \\
Leo~I                  & 10:08:27.5 & $+$12:18:21 & $-11.78$ & $-1.48$ & $ 12.1$ & 35,35,21,47 \\
Sculptor               & 01:00:04.4 & $-$33:43:07 & $-10.82$ & $-1.73$ & $  3.1$ & 35,35,21,47 \\
Antlia~II              & 09:35:13.9 & $-$36:41:56 & $ -9.86$ & $-1.90$ & \nodata & 18,18,18,NA \\
Leo~II                 & 11:13:27.0 & $+$22:09:10 & $ -9.74$ & $-1.68$ & $  7.4$ & 35,35,21,47 \\
Carina                 & 06:41:37.6 & $-$50:57:33 & $ -9.45$ & $-1.80$ & $ 11.6$ & 35,35,11,47 \\
Ursa~Minor             & 15:08:58.1 & $+$67:13:20 & $ -9.03$ & $-2.12$ & $  4.7$ & 35,35,21,47 \\
Sextans                & 10:13:03.1 & $-$01:36:48 & $ -8.94$ & $-1.97$ & $  1.8$ & 35,35,21,27 \\
Draco                  & 17:20:16.4 & $+$57:55:07 & $ -8.88$ & $-2.00$ & $  3.6$ & 35,35,21,47 \\
Canes~Venatici~I       & 13:28:02.2 & $+$33:33:08 & $ -8.73$ & $-1.91$ & $  5.5$ & 35,35,21,47 \\
Crater~II              & 11:49:14.4 & $-$18:24:47 & $ -8.20$ & $-2.16$ & $  3.3$ & 43,43,18,46 \\
Leo~T                  & 09:34:55.0 & $+$17:02:54 & $ -7.56$ & $-1.91$ & $ 12.2$ & 35,35,21,47 \\
Eridanus~II            & 03:44:20.1 & $-$43:32:02 & $ -7.10$ & $-2.38$ & $  2.0$ & 8,8,28,42 \\
Bootes~I               & 14:00:04.8 & $+$14:30:49 & $ -6.02$ & $-2.34$ & $  1.1$ & 35,35,15,2 \\
Hercules               & 16:31:05.3 & $+$12:47:07 & $ -5.83$ & $-2.47$ & $  1.8$ & 35,35,21,2 \\
Centaurus~I            & 12:38:20.4 & $-$40:54:07 & $ -5.55$ & \nodata & $  0.9$ & 32,32,NA,32 \\
Canes~Venatici~II      & 12:57:10.2 & $+$34:19:21 & $ -5.17$ & $-2.35$ & $  1.1$ & 35,35,21,2 \\
Ursa~Major~I           & 10:34:44.4 & $+$51:55:34 & $ -5.13$ & $-2.16$ & $  2.5$ & 37,35,21,2 \\
Leo~IV                 & 11:32:57.7 & $-$00:32:43 & $ -4.99$ & $-2.48$ & $  1.6$ & 35,35,15,2 \\
Hydra~II               & 12:21:42.0 & $-$31:59:10 & $ -4.86$ & $-2.02$ & $  2.2$ & 35,35,22,38 \\
Hydrus~I               & 02:29:33.4 & $-$79:18:32 & $ -4.71$ & $-2.52$ & \nodata & 25,25,25,NA \\
Eridanus~IV            & 05:05:45.1 & $-$09:30:54 & $ -4.70$ & \nodata & $  0.8$ & 4,4,NA,4 \\
Carina~II              & 07:36:25.6 & $-$57:59:57 & $ -4.50$ & $-2.44$ & \nodata & 45,45,29,NA \\
Ursa~Major~II          & 08:51:29.4 & $+$63:08:01 & $ -4.43$ & $-2.23$ & \nodata & 35,35,21,NA \\
Aquarius~II            & 22:33:55.5 & $-$09:19:39 & $ -4.36$ & $-2.30$ & \nodata & 44,44,44,NA \\
Indus~II               & 20:38:52.8 & $-$46:09:36 & $ -4.30$ & \nodata & \nodata & 9,9,NA,NA \\
Leo~V                  & 11:31:08.6 & $+$02:13:10 & $ -4.29$ & $-2.29$ & \nodata & 35,35,15,NA \\
Coma~Berenices         & 12:26:58.9 & $+$23:54:25 & $ -4.28$ & $-2.43$ & $  0.8$ & 35,35,21,2 \\
Pegasus~IV             & 21:54:09.4 & $+$26:37:12 & $ -4.25$ & $-2.67$ & $  1.3$ & 5,5,5,5 \\
Pisces~II              & 22:58:32.3 & $+$05:57:09 & $ -4.23$ & $-2.45$ & \nodata & 35,35,22,NA \\
Columba~I              & 05:31:25.7 & $-$28:02:33 & $ -4.20$ & \nodata & \nodata & 3,3,NA,NA \\
Pegasus~III            & 22:24:22.6 & $+$05:25:12 & $ -4.10$ & $-2.40$ & \nodata & 19,19,20,NA \\
Tucana~II              & 22:52:14.4 & $-$58:34:12 & $ -3.90$ & $-2.90$ & $  1.0$ & 1,1,6,38 \\
Grus~II                & 22:04:04.8 & $-$46:26:24 & $ -3.90$ & $-2.51$ & \nodata & 9,9,41,NA \\
Reticulum~II           & 03:35:40.9 & $-$54:03:05 & $ -3.88$ & $-2.65$ & $  1.5$ & 35,35,39,38 \\
Horologium~I           & 02:55:31.5 & $-$54:06:58 & $ -3.76$ & $-2.76$ & $  2.3$ & 35,35,24,38 \\
Pictor~I               & 04:43:47.8 & $-$50:17:07 & $ -3.67$ & \nodata & \nodata & 35,35,NA,NA \\
Tucana~IV              & 00:02:55.2 & $-$60:51:00 & $ -3.50$ & $-2.49$ & \nodata & 9,9,41,NA \\
Grus~I                 & 22:56:43.1 & $-$50:10:48 & $ -3.47$ & $-2.50$ & \nodata & 35,35,7,NA \\
Reticulum~III          & 03:45:26.4 & $-$60:27:00 & $ -3.30$ & \nodata & \nodata & 9,9,NA,NA \\
Pictor~II              & 06:44:43.2 & $-$59:53:49 & $ -3.20$ & \nodata & \nodata & 10,10,NA,NA \\
Bootes~II              & 13:58:03.4 & $+$12:51:19 & $ -2.94$ & $-2.79$ & \nodata & 35,35,16,NA \\
Willman~1              & 10:49:22.5 & $+$51:03:00 & $ -2.90$ & $-2.19$ & \nodata & 35,35,48,NA \\
Phoenix~II             & 23:39:58.3 & $-$54:24:18 & $ -2.70$ & \nodata & $  1.3$ & 36,36,NA,38 \\
Cetus~III              & 02:05:19.4 & $-$04:16:12 & $ -2.45$ & \nodata & \nodata & 14,14,NA,NA \\
Carina~III             & 07:38:31.2 & $-$57:53:59 & $ -2.40$ & $-2.65$ & \nodata & 45,45,17,NA \\
Segue~2                & 02:19:17.4 & $+$20:09:45 & $ -1.98$ & $-2.14$ & \nodata & 35,35,21,NA \\
Tucana~V               & 23:37:24.0 & $-$63:16:12 & $ -1.60$ & $-2.17$ & \nodata & 9,9,41,NA \\
Triangulum~II          & 02:13:18.0 & $+$36:10:13 & $ -1.60$ & $-2.24$ & $  0.9$ & 35,35,23,38 \\
Horologium~II          & 03:16:25.8 & $-$50:02:55 & $ -1.56$ & \nodata & \nodata & 35,35,NA,NA \\
Tucana~III             & 23:56:25.8 & $-$59:35:00 & $ -1.49$ & $-2.42$ & \nodata & 36,36,40,NA \\
Segue~1                & 10:07:00.1 & $+$16:04:32 & $ -1.30$ & $-2.71$ & \nodata & 35,35,12,NA \\
Draco~II               & 15:52:47.6 & $+$64:33:55 & $ -0.80$ & $-2.70$ & \nodata & 26,30,30,NA \\
Virgo~I                & 12:00:09.6 & $-$00:40:48 & $ -0.80$ & \nodata & \nodata & 13,13,NA,NA \\
Cetus~II               & 01:17:52.8 & $-$17:25:12 & $ +0.00$ & \nodata & \nodata & 9,9,NA,NA 
    \end{tabular}
    \caption{Milky Way Dwarf Galaxy Satellite Basic Properties, sorted by absolute magnitude $M_V$. Table based on \citet{Simon2019} with recent updates.
    Reference column indicates references for coordinates, $M_V$, [Fe/H], and $t_{\rm SF}$ separated by commas, with NA indicating no value for that data.
    The LMC and SMC are still star forming so $t_{\rm SF}$ is set to the age of the universe today.
    Reference numbers are 1: \citet{Bechtol2015}, 2: \citet{Brown2014}, 3: \citet{Carlin2017}, 4: \citet{Cerny2021}, 5: \citet{Cerny2022}, 6: \citet{Chiti2018}, 7: \citet{Chiti2022}, 8: \citet{Crnojevic2016}, 9: \citet{Drlica-Wagner2015}, 10: \citet{Drlica-Wagner2016}, 11: \citet{Fabrizio2012}, 12: \citet{Frebel2014}, 13: \citet{Homma2016}, 14: \citet{Homma2018}, 15: \citet{Jenkins2021}, 16: \citet{Ji2016b}, 17: \citet{Ji2020}, 18: \citet{Ji2021}, 19: \citet{Kim2015}, 20: \citet{Kim2016}, 21: \citet{Kirby2013}, 22: \citet{Kirby2015}, 23: \citet{Kirby2017}, 24: \citet{Koposov2015}, 25: \citet{Koposov2018}, 26: \citet{Laevens2015}, 27: \citet{Lee2009}, 28: \citet{Li2017}, 29: \citet{Li2018}, 30: \citet{Longeard2018}, 31: \citet{Majewski2003}, 32: \citet{Mau2020}, 33: \citet{McConnachie2012}, 34: \citet{Mucciarelli2017}, 35: \citet{Munoz2018}, 36: \citet{Mutlu-Pakdil2018}, 37: \citet{Okamoto2008}, 38: \citet{Sacchi2021}, 39: \citet{Simon2015}, 40: \citet{Simon2017}, 41: \citet{Simon2020}, 42: \citet{Simon2021}, 43: \citet{Torrealba2016}, 44: \citet{Torrealba2016b}, 45: \citet{Torrealba2018}, 46: \citet{Walker2019}, 47: \citet{Weisz2014}, 48: \citet{Willman2011}
}
    \label{tab:dwarfprop}
\end{table}

The most important parameter describing a dwarf galaxy is its stellar mass, which spans 7 orders of magnitude ($10^3 - 10^{10}\,M_\odot$). The stellar mass is inferred from a galaxy's observed luminosity which is typically given in absolute $V$-band magnitudes ($M_V$) and can be converted using
\begin{equation}\label{eq:mstarmv}
    M_\star = \eta\,10^{0.4(4.83 - M_V)}
\end{equation}
where $\eta$ is the mass-to-light ratio and is $\approx 2.2$ for old, metal-poor stellar populations  \citep{Ji2016b}. $\eta$ decreases to $1.0-1.5$ for younger stellar populations \citep{Kirby2013}.
The constant 4.83 is the absolute $V$ magnitude of the Sun.
Because of the large dynamic range, it is common to divide the dwarf galaxies into high- and low-mass galaxies, with a typical threshold that relatively luminous dwarf galaxies with $M_V < -7.7$ ($M_\star \gtrsim 2 \times 10^5\,M_\odot$) are called ``classical dwarf spheroidal'' (dSph) galaxies, while galaxies with luminosity $M_V > -7.7$ are deemed ``ultra-faint" dwarf galaxies (UFDs). 

\begin{figure}
\centering
    \includegraphics[width=0.7\linewidth]{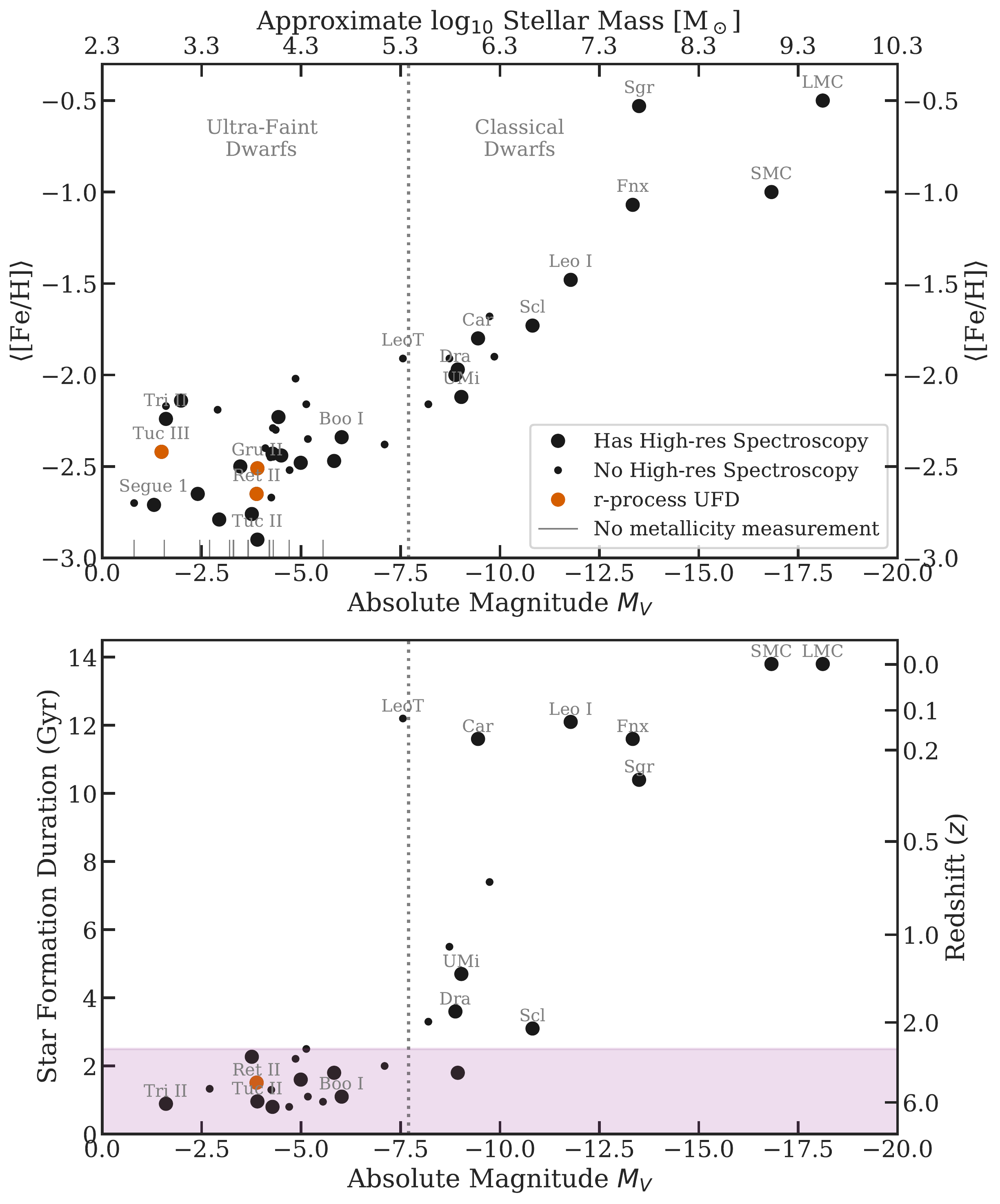}
    \caption{
    Summary of bulk dwarf galaxy properties.
    In both panels, the large black points indicate galaxies for which high-resolution spectroscopic data has been obtained, and thus the potential to measure abundances of the $r$-process elements. The small black points indicate galaxies with no high-resolution spectroscopy. The dotted black vertical line shows the division between Ultra-Faint Dwarf Galaxies and Classical Dwarf Galaxies.
    \emph{Top panel:} Absolute magnitude $M_V$ vs. mean metallicity $\langle\mbox{[Fe/H]}\rangle$. The top axis indicates the approximate stellar mass corresponding to each magnitude, assuming a mass-to-light ratio of 2.2 in Equation~\ref{eq:mstarmv}. This overestimates the mass of high luminosity galaxies, where the mass-to-light ratio can decrease to $1.0-1.6$ \citep{Kirby2013}.
    In general, more massive dwarf galaxies have higher median metallicities.
    \emph{Bottom panel:} Absolute magnitude $M_V$ vs the star formation duration (in Gyr).
    The star formation duration is defined to be the time to form ${\approx}90\%$ of a galaxy's stars.
    The shaded pink region at the bottom indicates the range of ages where current galaxy age determinations are not sufficiently precise to robustly distinguish between different ages, i.e. galaxies in that range should be thought of as having essentially identical star formation durations.
    Galaxies with stellar masses below ${\approx}10^5 M_\odot$ uniformly formed in the first 1-2 billion years, while more massive galaxies are able to form stars for longer times.
    The original references for data in this figure are given in Table~\ref{tab:dwarfprop}.
    }
    \label{fig:dwarfprop}
\end{figure}

The top panel of Figure~\ref{fig:dwarfprop} shows the galaxy ``luminosity-metallicity relation'' (often abbreviated LZR, where Z is a common symbol representing metallicity; or MZR for the ``mass-metallicity relation''). The y-axis $\langle\mbox{[Fe/H]}\rangle$ is the mean of individual star $\mbox{[Fe/H]}$ measurements, so it is best thought of as the median metallicity of the whole galaxy (due to the $\log_{10}$ in the definition of [Fe/H]).
It is clear that there is an overall scaling where more massive galaxies have higher median metallicities.
This relation extends continuously to higher mass galaxies and appears to be present independent of a galaxy's large-scale environment \citep{Kirby2013}.
However, it is important to remember that metal-poor stars are in fact present in all galaxies, though they are a small fraction of the stars in metal-rich galaxies \citep[e.g.,][]{Frebel2010a,Chiti2018,Reggiani2021}.

In the Figure,  galaxies with high-resolution spectroscopic abundances are highlighted as larger black points. These are the only galaxies in which some $r$-process element abundances can be measured. Only about 50\% of the known Milky Way dwarf galaxies have been studied with high-resolution spectroscopy, though current studies span the entire range of masses and metallicities.
Three of the ultra-faint dwarf galaxies (Ret~II, Tuc~III, Gru~II) are highlighted in red as $r$-process ultra-faint dwarf galaxies, which will be discussed in Section~\ref{sec:ufd}.

The bottom panel of Figure~\ref{fig:dwarfprop} shows the primary reason to divide galaxies by mass which corresponds to their fundamentally different formation histories.
The $y-$axis of that panel is the star formation duration, defined as the time it takes for the galaxy to form ${\approx}90\%$ of its present-day stellar mass.
Low-mass dwarf galaxies occupy low-mass dark matter halos with present-day total masses of $M_{\rm halo} \lesssim 10^9\,M_\odot$. These halos are not massive enough to accrete gas from the intergalactic medium after cosmic reionization occurred at $z \approx 6$. Thus, the ultra-faint dwarf galaxies have their star formation halted by reionization, and their star formation histories appear to uniformly terminate somewhere between $z=2-6$, corresponding to 1-2\,Gyr after the Big Bang \citep{Weisz2014,Brown2014, RodriguezWimberly2019}.
In contrast, more massive galaxies can continue to accrete gas and form stars after $z=6$, likely only quenching when they interact with a more massive galaxy \citep{Geha2012,Fillingham2019}. Of the Milky Way's known satellite galaxy population (i.e. within its 300 kpc dark matter halo virial radius), only the Large and Small Magellanic Clouds (the most massive satellites and orbiting for only 1-2\,Gyr around the Milky Way, \citealt{Besla2007, Kallivayalil2013}) have retained any gas and are still forming stars.
However, even low-mass dwarf galaxies outside the virial radius (e.g., the distant dwarf galaxy Leo T at 400 kpc away) can still retain gas and actively form stars.

\subsection{Chemical Enrichment and Evolution of Non-$r$-process Elements in Dwarf Galaxies}

Before describing the observed $r$-process signatures in dwarf galaxies in Section~\ref{sec:dwarfrproc}, first consider a brief overview of the chemical evolution in dwarf galaxies.

Chemical element abundances are best derived from high-resolution spectrum, but only half of the ${\sim}60$ dwarf galaxy satellites of the Milky Way have even a single star observed with the high-resolution spectroscopy. 
The elements measurable with stellar spectroscopy are generally grouped into four categories: The $\alpha$-elements (e.g., O, Mg, Ca, Ti) that are made predominantly in core-collapse supernovae; the iron-peak elements (e.g., Mn, Fe, Ni, Zn) that are produced predominantly in thermonuclear (Type Ia) supernovae; the $s$-process elements that are synthesized predominantly in evolved, low and intermediate mass Asymptotic Giant Branch (AGB) stars; and the $r$-process elements whose origin remains uncertain.
Recall that element sources occurring shortly after a burst of star formation are regarded ``prompt'' (e.g., core-collapse supernovae following the deaths of massive stars), and element sources occurring longer after a burst of star formation are ``delayed'' (e.g., Type Ia SNe, low mass AGB stars, and neutron star mergers). 

The first and most important feature of dwarf galaxy chemical evolution is the fact that metallicity increases at different rates in different dwarf galaxies.
Gas in lower-mass galaxies increases its metallicity at a slower rate than gas in higher-mass galaxies, for two reasons. First, the lower gravitational potential well of low-mass galaxies means that it is easier to directly expel metals out of low mass dwarf galaxies \citep[e.g.,][]{Dekel2003,Kirby11b,Brauer2021}.
Second, stellar feedback is more efficient at preventing star formation in low-mass galaxies, so the overall star formation efficiency, and thus metal production rate, is lower \citep[e.g.,][]{Ishimaru15,Ji2016b}.
Together, these effects mean that the expected age of a metal-poor star depends strongly on the galactic environment that it forms in. Metallicities can thus reasonably be interpreted as relative ages in dwarf galaxies, but this does not applied to stars or stellar samples from the Milky Way stellar halo.

\begin{figure}
    \centering
    \includegraphics[width=0.95\linewidth]{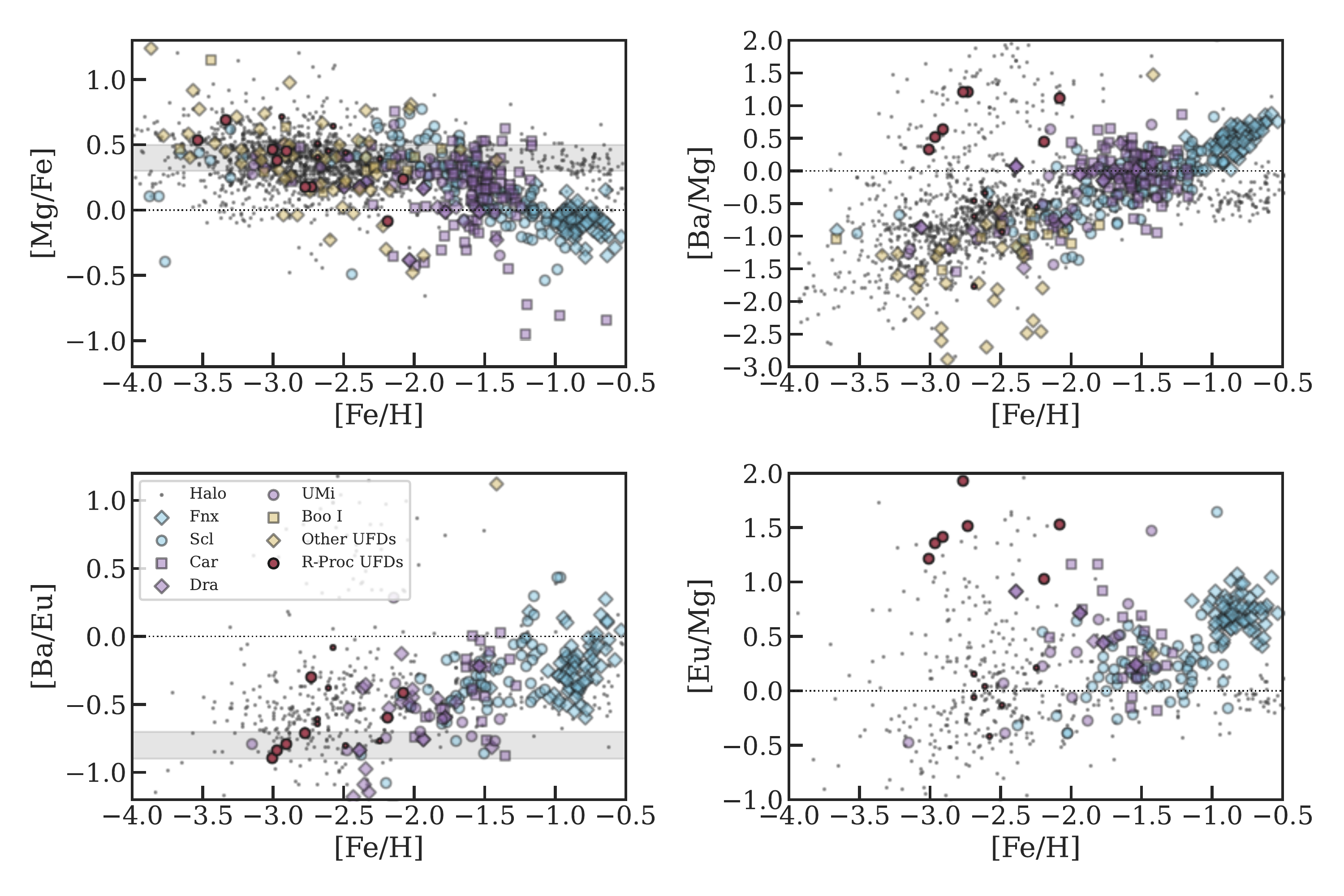}
    \caption{Chemical evolution in dwarf galaxies.
    Each point represents the chemical abundances of one star, with colors and shapes indicating the formation site of a star. Data with upper limits have not been plotted.
    Grey points are stars in the Milky Way's stellar halo \citep{Abohalima18}.
    Light blue points are stars from the relatively massive $10^6-10^7 M_\odot$ Fornax and Sculptor dSph galaxies \citep{Letarte2010,Hill2019}.
    Purple points are stars from the lower mass $10^6 M_\odot$ Carina \citep{Venn2012,Lemasle2012,Norris2017}, Draco \citep{Cohen2009,Tsujimoto2017}, and Ursa Minor \citep{Cohen2010,Kirby2012,Ural2015} dSph galaxies.
    Yellow points are stars from ultra-faint dwarf galaxies (UFDs, $<10^5 M_\odot$), using the literature compilation in \citealt{Ji2020}) displaying low neutron-capture element abundances. Bootes I is plotted separately as it shows a rise in $s$-process at high metallicity, contrasting with other UFDs \citep{Frebel16}.
    Red points are stars from $r$-process enhanced UFDs, where larger points are from the highly $r$-process enhanced galaxy Reticulum~II and smaller points are the other two UFDs \citep{Ji2016c,Marshall2019,Hansen2021}.
    \emph{Top left:} [Mg/Fe] vs [Fe/H], which indicates the relative enrichment of Type Ia and core-collapse supernovae. An approximate plateau at $\mbox{[Mg/Fe]} \sim 0.4$ exists at low metallicity for all systems (shown as shaded grey bar), but transitions to lower [Mg/Fe] at an [Fe/H] that broadly depends on galaxy mass.
    \emph{Bottom left:} [Ba/Eu] vs [Fe/H], which indicates the relative amount of $s$-process vs $r$-process enhancement. The shaded grey bar shows $\mbox{[Ba/Eu]} \sim -0.8$, which is the universal value found in the solar $r$-process residual and in $r$-process enhanced stars. At higher metallicities, the onset of delayed $s$-process enrichment from AGB stars adds a lot of barium relative to europium, resulting in a gradual increase in [Ba/Eu] with metallicity. There is one UFD star with high [Ba/Eu] that is the result of $s$-process from AGB mass transfer.
    \emph{Top right:} [Ba/Mg] vs [Fe/H]. The origin of Ba depends on metallicity and environment. At high metallicities, [Ba/Mg] shows the rise of the $s$-process by comparing delayed AGB to prompt core-collapse supernovae enrichment. At low metallicities, Ba is predominantly from $r$-process, where the UFDs show a very large scatter due to stochastic early $r$-process enrichment.
    Note that all stars with sufficient data quality to detect $\mbox{[Ba/H]} > -4$ have successfully detected Ba \citep{Roederer13}.
    \emph{Bottom right:} [Eu/Mg] vs [Fe/H], which compares the relative delay times of the $r$-process source compared to core-collapse supernovae.
    It is clear that dwarf galaxies of different masses have different overall [Eu/Mg] (especially comparing Fornax to Sculptor), which is strong evidence for the existence of delayed $r$-process enrichment.
    }
    \label{fig:chemevol}
\end{figure}

The next most important feature of chemical evolution is the so-called ``$\alpha$-plateau and knee'' \citep{Tinsley80,Matteucci1990,Tolstoy2009}, shown in the top-left panel of Figure~\ref{fig:chemevol}.
These panels compare an $\alpha$-element like Mg, produced only in core-collapse supernovae, to an iron-peak element such as Fe, produced in both core-collapse supernovae and Type Ia supernovae.
In every galaxy, the low-metallicity stars (those forming earlier in a dwarf galaxy's life) tend to have enhanced ratios of [Mg/Fe] $\gtrsim 0.3$ (i.e., more Mg than Fe relative to the solar composition for which the ratio is zero).
At some metallicity, that ratio begins to decrease, and the more metal-rich stars (those forming later in a dwarf galaxy's evolution) end up with progressively lower ratios of [Mg/Fe] all the way to ${\lesssim} 0$.
The interpretation here is that early in a galaxy's history, it is predominantly enriched by core-collapse supernovae which produce a large amount of $\alpha$-elements compared to the solar composition.
At some point, the delayed Type Ia supernovae begin to dominate, synthesizing enough Fe to reduce the [$\alpha$/Fe] ratio, and thus producing a ``kink'' or ``knee'' in the [$\alpha$/Fe] vs. [Fe/H] chemical abundance trend.
Because the timing of the onset of Type Ia supernovae is a property of the underlying stellar populations, galaxies with efficient metal production will have their $\alpha$-knee located at a \emph{higher} metallicity, while galaxies with inefficient metal production will have their $\alpha$-knee located at a \emph{lower} metallicity.
Since galaxies of higher mass tend to produce stars and metals more efficiently, the metallicity location of the $\alpha$-knee is thus positively correlated with stellar mass. One major exception is galaxies that form so quickly that they never experience enrichment by Type~Ia supernovae, leaving a flat [$\alpha$/Fe] trend over their entire metallicity range \citep{frebel12}. So far, only the Segue~1 ultra-faint dwarf galaxy displays this signature \citep{Frebel2014}.

Note that the above simple picture explains the overall evolution in most galaxies and the overall trend with galaxy mass, but it is not correct in detail.
In particular, a common misconception is that the $\alpha$-knee indicates the exact time that the first Type~Ia supernovae ``turn on'' (e.g., ${\approx}100$\,Myr after the first burst of star formation, \citealt{maoz12}).
However, in fact it occurs when Type~Ia supernovae ejecta \emph{dominate} over core-collapse supernovae, which can happen much later \citep[e.g.,][]{Maoz2017}. \citet{Theler2020} and \citet{Kirby2020} have recently both advocated that the \emph{slope} of the [$\alpha$/Fe] vs. [Fe/H] decline is a more meaningful tracer indicator than the knee location.
This would naturally explain why the Large and Small Magellanic Clouds, the most massive dwarf galaxies around the Milky Way, have some of the lowest metallicity $\alpha$-knees \citep{Nidever2020}.

A third important feature is the ``rise of the $s$-process'' \citep[e.g.,][]{Simmerer04,Venn04}.
The bottom-left panel of Figure~\ref{fig:chemevol} shows the [Ba/Eu] ratio vs metallicity [Fe/H].
Assuming the solar $r$-process pattern is universal, the $r$-process intrinsically produces a fixed ratio of [Ba/Eu] = $-0.9$ to $-0.7$ \citep{Sneden2008}.
It is empirically observed that the most metal-poor stars in a galaxy cluster around this intrinsic value, while the more metal-rich stars have higher [Ba/Eu], which is interpreted as delayed $s$-process enrichment from AGB stars.
The different [Ba/Y] ratio in dwarf galaxies is interpreted as being due to enrichment by the $s$-process operating in metal-poor AGB stars where the metal-poor $s$-process produces less Y than Ba compared to metal-rich AGB stars, due to the lower neutron-to-seed ratio in metal-poor AGB stars \citep[e.g.,][]{Venn04,Lugaro12} leading to a larger relative production of the heavier elements.

Finally, an important distinction between chemical \emph{enrichment} and chemical \emph{evolution} in dwarf galaxies needs to be made.
Chemical enrichment is the phase where metal production and observed stellar abundances can be linked to discrete enrichment sites (e.g., individual core-collapse supernovae).
Chemical evolution is the long-term average evolution of the galaxy composition, once it is no longer possible to distinguish between individual sources of elements.
For most dwarf galaxies, the chemical enrichment phase is restricted to the earliest times and as such only reflected in the abundances of the most metal-poor stars in the system.
However, the lowest mass dwarf galaxies may never form enough stars to exit the chemical enrichment phase.
This can be seen in Figure~\ref{fig:chemevol} on all panels, where the scatter in metal ratios at low [Fe/H] is generally larger than that at higher [Fe/H].

\section{$R$-process Signatures in Dwarf Galaxies}\label{sec:dwarfrproc}

This section gives a general summary of all existing $r$-process chemical evolution trends in dwarf galaxies, as well as specifics about  important individual dwarf galaxies.
Dwarf galaxies are grouped into three categories: The ultra-faint dwarf galaxies which are dominated by stochastic $r$-process enrichment (Section~\ref{sec:ufd}); the large intact dwarf galaxies which display a transition from $r$-process enrichment to $r$-process element evolution (Section~\ref{sec:cldw}); and the tidally disrupted dwarf galaxies now observable as stellar streams and kinematic structures in the stellar halo (Section~\ref{sec:disrupteddwarfs}).

\subsection{Stochastic $R$-process Events in Ultra-Faint Dwarf Galaxies}\label{sec:ufd}

Ultra-faint dwarf galaxies typically have masses of $M_\star \lesssim 2 \times 10^5 M_\odot$; systems more massive that this are the classical dwarf spheroidal galaxies. Note that this dividing mass is somewhat arbitrary and more reflects the history of their relatively recent discoveries than any underlying physical properties. However, $2 \times 10^5 M_\odot$ roughly corresponds to galaxies insufficiently massive to form stars after reionization at $z \sim 6$ \citep{Simon2019}.
There are ${\sim}50$ known ultra-faint dwarf galaxies, of which ${\sim}20$ have high-resolution spectroscopy available of at least one star and in which neutron-capture elements could in principle be detected.

The vast majority of ultra-faint dwarf galaxies display unusually low neutron-capture elements -- so low that it is not even possible to identify whether the neutron-capture source is $r$-process or $s$-process as only Sr and Ba can be measured \citep{Frebel2015,Ji2019}.
The fundamental explanation is the low stellar mass of ultra-faint dwarf galaxies. For example, if the average rate of $r$-process production is about one $r$-process event per ${\sim}$1,000 core-collapse supernovae, it takes ${\sim}10^5 M_\odot$ of stellar mass formed in order to produce one single $r$-process event.
Thus, individual ultra-faint dwarf galaxies are unlikely to display evidence of $r$-process enrichment; but as a population, a few UFDs should be $r$-process-rich galaxies.

\begin{figure}
\centering
    \includegraphics[width=\linewidth]{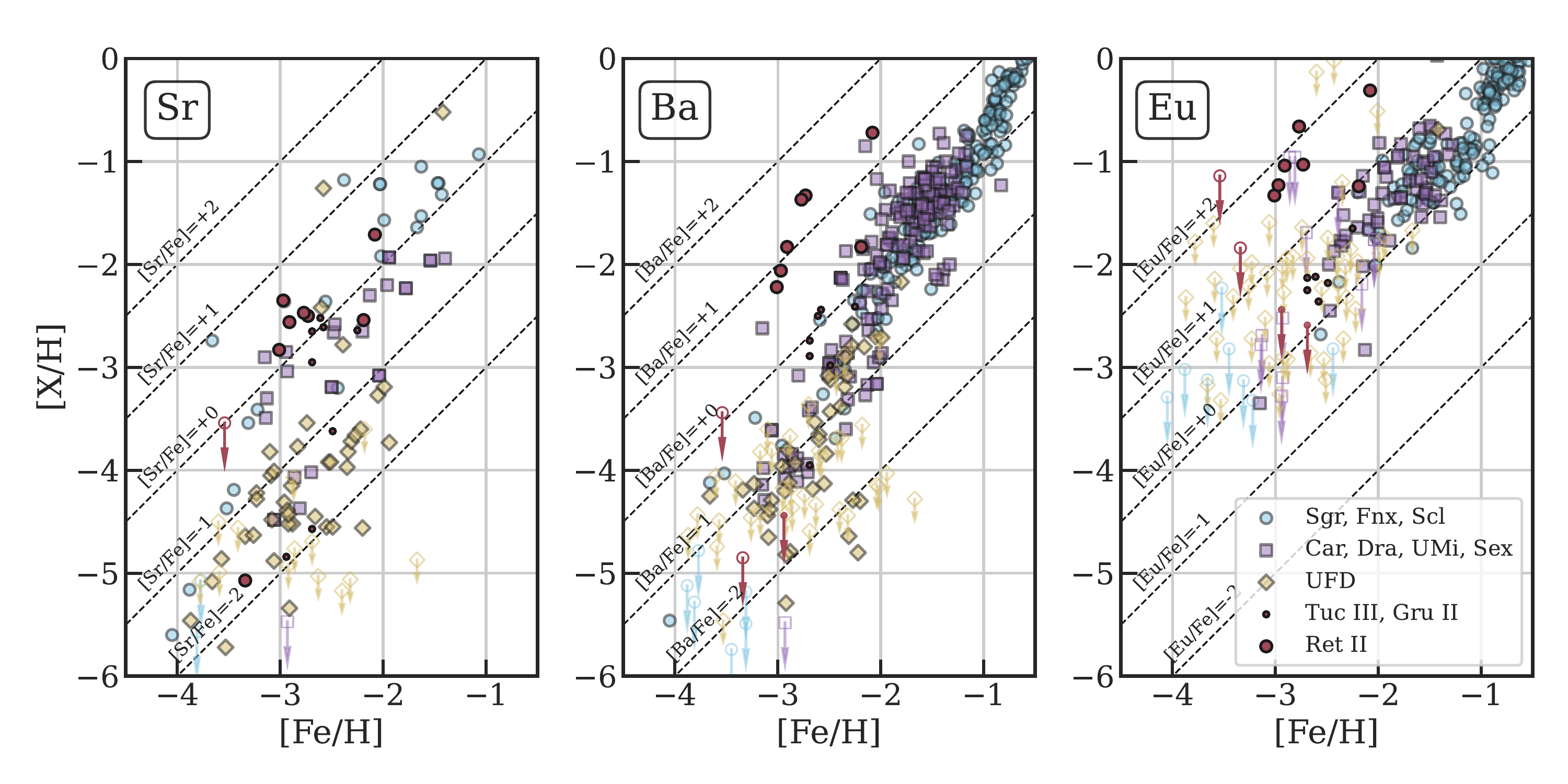}
    \caption{Neutron-capture elements in dwarf galaxies depending on their stellar mass.
    Blue circles are relatively massive classical dSphs, purple squares are intermediate mass classical dSphs, yellow diamonds are ultra-faint dwarf galaxies with no indication of $r$-process elements, red circles are the three ultra-faint dwarf galaxies with some enhancement in $r$-process elements.
    Ret~II is shown as larger symbols to emphasize its higher [Eu/Fe] compared to Tuc~III and Gru~II.
    Open symbols with downward pointing arrows are upper limits, i.e. no detection of that element was possible.
    Most ultra-faint dwarf galaxies have low neutron-capture abundances compared to classical dSphs, including in the range $-3 < \mbox{[Fe/H]} < -2$, except for the rare $r$-process enhanced ultra-faint dwarf galaxies.
    Note that the upper limits for Sr and Ba are quantitatively meaningful in many cases, but the upper limits for the $r$-process tracer Eu just indicate that Eu is difficult to detect at low abundances.
    The increase in Ba for classical dSphs and a few ultra-faint dwarf galaxies at [Fe/H] $\gtrsim -2$ indicates the onset of significant $s$-process enrichment.
    \label{fig:ufdncap}}
\end{figure}

This line of reasoning was brought about by the discovery that the ultra-faint dwarf galaxy Reticulum~II is $r$-process enhanced, with over 2/3 of its stars having $\mbox{[Eu/Fe]} \sim 1.7$ (Ret~II; \citealt{Ji2016b,Roederer2016}; see Figure~\ref{fig:ufdncap} for updated data). Ret~II has $r$-process abundances over $10-100{\times}$ higher than what is found in other ultra-faint dwarf galaxies, indicating that it preserves the chemical abundance signature of a single $r$-process event. By considering the 10 ultra-faint dwarf galaxies available at that time, \citet{Ji2016b} derived strong $r$-process production constraints from inferring that the event produced $10^{-5.5}-10^{-3.5} M_\odot$ of europium, and that the relative rate of $r$-process events across the ultra-faint dwarf population was 1 $r$-process event for every 1,000-2,000 core-collapse supernovae (also see \citealt{Beniamini2016b}). These properties were consistent with theoretical expectations for neutron star mergers \citep{Goriely11,Dominik12}, and the theoretical values have since been validated by multi-messenger observations of the neutron star merger GW170817 and its kilonova \citep[e.g.,][]{Cote2018}. An immediate major concern with neutron star mergers is the fact that neutron star binaries typically get a velocity kick upon formation, which would cause them to be unbound from such a low-mass galaxy, in which case the $r$-process elements would not be captured by later generations of star formation \citep{Bramante2016,Bonetti2019}. This scenario can be avoided if the binary merges sufficiently quickly, and it is an open question whether binary neutron stars can merge sufficiently quickly to deposit their $r$-process elements in ultra-faint dwarf galaxies \citep{Beniamini2016,Safarzadeh2019}.
An alternate likely possibility is that a rare core-collapse supernovae would produce a large amount of $r$-process, perhaps through strong magnetorotationally driven jets or disk winds from collapsars (see Section~\ref{candidates}).

The most recent study of Ret~II finds that 70\% of its stars are $r$-process enhanced, and in those stars the $r$-process material is very homogeneously distributed \citep{Ji2022}.
Current theoretical expectations for galaxies of Ret~II's mass are that the stars without $r$-process enhancement should be born in the same galaxy rather than accreted \citep{Griffen2018}.
Based on metal mixing simulations, this suggests that ${\gtrsim}100$\,Myr must have elapsed between the \emph{production} of $r$-process material and the formation of the first 30\% of Ret~II stars \citep{Tarumi2020}. This suggests a very prompt source of $r$-process elements and a relatively large $r$-process yield of $M_r \gtrsim 10^{-1.5} M_\odot$.

Since the discovery of Ret~II, two other ultra-faint dwarf galaxies or galaxy candidates have been clearly shown to contain $r$-process enhancements: Tucana~III \citep{Hansen2017} and Grus~II \citep{Hansen2020}.
Unlike Ret~II, these two systems have moderate $r$-process enhancements [Eu/Fe] $\sim +0.5$, which can either be explained by dilution into different gas mass reservoirs \citep{Safarzadeh2017,Tarumi2020} or variable $r$-process yields \citep{Ji2019b}.
Gru~II is clearly a bona fide dwarf galaxy, and like Ret~II it shows a transition from metal-poor $r$-process-free stars to (relatively) metal-rich rich $r$-process-enhanced stars.
In contrast, there is still debate whether Tuc~III is a ultra-faint dwarf galaxy or a globular cluster, though the current evidence suggests it is more likely an ultra-faint dwarf galaxy \citep{Marshall2019}.
There have not yet been more detailed studies using all galaxies that could result in more precise (but more model-dependent) determinations of the $r$-process yield and rate from ultra-faint dwarf galaxies.

An important question is whether all the $r$-process enhanced stars in the Milky Way halo could be attributed to tidally disrupted UFDs \citep{Tsujimoto14b,Ishimaru15,Ji2016c,Ojima2018}.
\citet{Brauer2019} modeled the assembly of a Milky Way stellar halo, finding that typically 50\% of the metal-poor $r$-II stars originate from UFDs, though this fraction varies from 20\% to 100\% depending on the specific Milky Way accretion history. Other, similar studies find fractions close to 100\% as well \citep{hirai22}. The other $r$-II stars thus likely originate from more massive dwarf galaxies, such as the $r$-process enhanced stars found in Ursa Minor \citep{Cohen2010} or Fornax \citep{Reichert2021}.

\subsection{The Rise of the $R$-process: From Enrichment to Evolution}\label{sec:cldw}

Galaxies with stellar mass exceeding ${\sim}10^6 M_\odot$ will experience $\gtrsim 10$ $r$-process enrichment events over the course of their lives.
The larger number of $r$-process events induces a transition from \emph{chemical enrichment} to \emph{chemical evolution}: Rather than being dominated by individual stochastic $r$-process enrichment events, instead the galaxy's $r$-process composition can reasonably be described as continuous chemical evolution.
To estimate how many $r$-process events are required for this transition to occur, note that an unresolvable $r$-process abundance scatter would be 0.1-0.2\,dex scatter (a typical systematic uncertainty on [Eu/Fe] measurements in dwarf galaxies). This translates to 25\%-60\% relative $r$-process scatter, or equivalently the Poisson scatter of $r$-process yields from 3-15 $r$-process events.

The bottom right panel of Figure~\ref{fig:chemevol} shows $r$-process chemical evolution in dwarf galaxies as measured by [Eu/Mg] vs [Fe/H]. Each point on the plot indicates the chemical abundances of a single star in a dwarf galaxy, and the points have been color-coded according to the typical mass of the galaxy (with different symbols representing individual dwarf galaxies).
Traditionally, the literature has focused on the [Eu/Fe] vs [Fe/H] evolution plane, because Fe is the easiest-to-measure and most precise stellar metallicity indicator. However, it was emphasized in \citet{Tolstoy2009} and more recently by \citet{Skuladottir2019} that Fe may not be a simple reference element to study dwarf galaxy chemical evolution, because it is produced in both core-collapse supernovae and Type Ia supernovae.
An empirical way to address this is using Mg instead of Fe as a reference element, as Mg is only produced in core-collapse supernovae. It can also be addressed with chemical evolution modeling, e.g. by simultaneously fitting [Mg/Fe] observations \citep[e.g.,][]{Molero2021,delosReyes2022} or calculating the Ia contribution with [Mg/Fe] \citep[e.g.,][]{Kirby2019}.

The dwarf galaxies Draco, Ursa Minor, Sextans, and Carina have present-day stellar masses $\sim 10^6 M_\odot$, so they should display a transition from discrete $r$-process enrichment at the low-metallicity end to continuous $r$-process trends at high-metallicity \citep{Duggan2018,Skuladottir2019}.
Draco is the clearest example of this, as it shows multiple distinct [Eu/H] plateaus that may correspond to discrete $r$-process enrichment events, though its neutron-capture elements are dominated by $s$-process enrichment \citep{Cohen2009,Tsujimoto2015,Tsujimoto2017}.
Ursa Minor is dominated by $r$-process enrichment, but it has unusual Eu trends and large scatter that likely suggest it is still dominated by stochastic $r$-process enrichment \citep{Cohen2010,McWilliam2018}.
Sextans unfortunately currently has only limited $r$-process measurements \citep{Theler2020, Reichert2020}, but at the high-metallicity end it currently appears to have low Eu scatter indicating a transition to continuous chemical evolution.
Carina is a strange case as its star formation history has at least three distinct bursts \citep{Hurley-Keller1998,Weisz2014}.
Its early history seems to have clear indications of inhomogeneous $r$-process abundances \citep{Venn2012}, and its later history has more continuous Eu evolution \citep{Lemasle2012, Norris2017}.

Sculptor, Fornax, and Sagittarius are sufficiently massive that they should be dominated by chemical evolution rather than enrichment. Sculptor is one of the best-studied dwarf galaxies with a very simple star formation history and smooth $r$-process chemical evolution \citep{Duggan2018, Hill2019, Skuladottir2019, Molero2021, delosReyes2022}. There is still debate about the relative importance of prompt vs delayed $r$-process sources in Sculptor, but it clearly has reached the point of continuous chemical evolution with [Eu/Mg] ${\approx}0$.
Fornax \citep{Letarte2010, Lemasle2014} and Sagittarius \citep{Bonifacio2000,McWilliam2013,Hansen2018} are two of the more massive dwarf spheroidal galaxies with $r$-process elements, both with enhanced [Eu/Mg] $\sim 0.4$. These indicate that Fornax and Sagittarius have a relatively large contribution from delayed sources of $r$-process elements compared to Sculptor, which matches the difference in star formation durations seen in the bottom panel of Figure~\ref{fig:dwarfprop}.
Note that there is clearly some inhomogeneous $r$-process enrichment in Fornax, as some of its most metal-rich stars have extreme Eu enhancements \citep{Reichert2021}.

The Large and Small Magellanic Clouds (LMC, SMC) are the only gas-rich and actively star-forming dwarf galaxy satellites of the Milky Way.
The most comprehensive recent study of the LMC and SMC is by \citet{Nidever2020}. They were unable to study $r$-process chemical evolution, but they did show that the LMC and SMC have an inefficient star formation history, such that it should be dominated by delayed-enrichment sources even at relatively low metallicity.
This is consistent with the findings of \citet{Reggiani2021}, who measured $r$-process abundances of metal-poor stars ($-2.5 \lesssim \mbox{[Fe/H]} \lesssim -1.5$) in the LMC and SMC, finding they were unusually enhanced and thus favoring delayed sources of $r$-process production.
At higher metallicities, there is also evidence of enhanced [Eu/Mg] implying delayed $r$-process enrichment \citep{Naidu2022}, though also higher levels of contamination by the $s$-process in AGB stars \citep{VanderSwaelmen2013}.
Unlike the simpler dSph galaxies, the LMC and SMC have distinct physical components (e.g. disk, bar, halo) with their own formation histories, and thus it is wise to separate chemical abundance interpretations based on the component being studied \citep[e.g.,][]{Pompeia2008, VanderSwaelmen2013,Reggiani2021}.

\subsection{Tidally Disrupted Dwarf Galaxies: Stellar Streams and Dissolved Galaxies}\label{sec:disrupteddwarfs}

Dwarf galaxies that accreted very early on into the Milky Way will be tidally disrupted by the Milky Way's gravitational potential. Their stars will be strewn throughout the Milky Way's stellar halo, but the stars will still retain their chemical compositions and some conserved kinematic quantities (e.g. energy, angular momentum, and actions).
The advent of all-sky astrometry from the \emph{Gaia} space mission \citep{GaiaCollaboration2016} has recently enabled detailed calculations of the kinematics, and has ushered in a new era of discovery and chemically characterization of such tidally disrupted dwarf galaxies. These galaxies can differ from the intact galaxies discussed previously, because they stop forming stars when they interact with the Milky Way \citep[e.g.,][]{Johnston2008,Fillingham2019}.

The task of breaking apart the stellar halo into discrete accreted dwarf galaxies is still ongoing, but there should only be a few massive dwarf galaxy accretions \citep[e.g.,][]{Deason2016,Brauer2019}. The broad picture of the massive Milky Way mergers is now largely in place \citep[e.g.,][]{Naidu2020,Kruijssen2020}.
The most significant dwarf galaxy merger is Gaia-Sausage-Enceladus \citep[or GSE,][]{Belokurov2018, Helmi2018}, which makes up over 50\% of the halo stars in the solar neighborhood (though a smaller fraction of the metal-poor stars).
This galaxy had a stellar mass of ${\approx}10^8 M_\odot$ when it merged into the Milky Way $10-12$ billion years ago. Several studies have shown that the metal-rich GSE stars have a high [Eu/Mg], indicating a preference for enrichment by delayed $r$-process sources \citep{Matsuno2021,Aguado2021,Naidu2022}.
Another significant merger is the Kraken or Heracles galaxy \citep{Kruijssen2020,Horta2021}, which is confined to the inner 5 kpc of the Milky Way and thus merged with the Milky Way before GSE. \citet{Naidu2022} showed that Kraken stars display a low [Eu/Mg] enhancement, so it may be dominated by prompt $r$-process sources.
A third likely dwarf galaxy merger is the Sequoia galaxy \citep{Matsuno2019,Myeong2019}, which is substantially more metal-poor and chemically distinct \citep{Matsuno2022} but does not yet have detailed $r$-process studies.
There should be many more tidally disrupted low-mass dwarf galaxies that can be identified in the Milky Way, and searches are ongoing to discover and characterize their nature \citep[e.g.,][]{Roederer2018,Naidu2020,Yuan2020,Gudin2021,Shank2022,Brauer2022, Mardini2022}.

Dwarf galaxies accreted more recently onto the Milky Way can be found as tidally unbound but still spatially coherent structures known as stellar streams.
Streams are very low surface brightness, and they are discovered through a combination of deep imaging and Gaia astrometry \citep[e.g.,][]{Belokurov2006,Grillmair2009,Shipp2018,Ibata2019}.
The most spectacular stellar stream is the Sagittarius stream, which is caused by the active tidal disruption of the Sagittarius dwarf galaxy \citep{Majewski2004}; but many stellar streams have no apparent progenitor system.
Streams with unknown progenitors require spectroscopic followup to confirm whether they are dwarf galaxies or globular clusters \citep[e.g.,][]{Li2022}.
There are now several spectroscopic studies of stellar streams that are producing  $r$-process constraints \citep[e.g.,][]{Roederer10,Casey2014,Ji2020b,Hansen2021,gull21,Limberg2021}.
Currently the data are still limited, but there are already as many dwarf galaxy streams as intact dwarf galaxies in the intermediate mass range of $10^5-10^6 M_\odot$, and it is likely many more will be discovered in the future.

The immense promise of the tidally disrupted dwarf galaxies for the $r$-process is that they could be the best of both worlds, containing stars close enough to obtain extremely high quality spectra for full $r$-process patterns while also having a galactic environment to interpret \citep[e.g.,][]{Naidu2022}.
However, a general challenge with interpreting the $r$-process content of tidally disrupted systems is ensuring the purity of stars studied \citep[e.g.,][]{Limberg2021}. Unlike intact dwarf galaxies, where a spatial cut clearly selects galaxy member stars, a tidally disrupted system can overlap with multiple other structures in position and velocity space. Membership in these systems must be probabilistic. It is thus difficult to study intrinsic scatter in abundances in tidally disrupted systems, but it is still useful to study mean chemical abundance trends.

\section{State of the Art and Path Ahead}\label{sec:conclusion}

The many discoveries and observations of $r$-process signatures in halo and dwarf galaxies confirm the existence of the $r$-process within early star forming environments as well as within dwarf galaxies at various times since the Big Bang. Still, information remains limited with regards to details on the astrophysical site and environment as well as other parameters that describe the operation of this nucleosynthesis process. 

The following open questions will eventually need to be answered to fully understand the origin of $r$-process elements:  

\begin{itemize}
\item How many astrophysical sites can be identified? If more than one, what is their relative importance in different environments and over cosmic time?
\item What is the average rate of $r$-process events? When did the first $r$-process events occur? How quickly does $r$-process production occur after star formation?
\item What is the $r$-process yield? Is it variable per site? How do yields for light and heavy $r$-process elements, as well as actinides, compare? 
\item How much does $r$-process material get diluted and how quickly does it reincorporated into stars? What is the explosion energy associated with $r$-process events? How does this depend on galactic environment?
\end{itemize}

Current data have begun to provide answers while future photometric and spectroscopic sky surveys as well as new instruments and telescopes will hopefully provide additional insight over the coming decades. The current state of the art is laid out in this section, which offers a glimpse of what is to come in terms of new data and telescopes.

\subsection{Observational Constraints on the $R$-process}

Despite the above questions about the nature and origin of the $r$-process, the body of data has, over the last decade, finally begun to reveal a number of important findings, constraints, anecdotes, and partial answers. 
Table~\ref{summarytable} summarizes these results, which represents the state of the art of observational results regarding the astrophysical $r$-process from metal-poor halo stars and dwarf galaxies.
Overall, it is clear that compositional information is best derived from the closer halo stars, while questions relying on the galactic environment benefit most from the dwarf galaxies.

\begin{table}[!h]
\tabcolsep7.5pt
\begin{center}
\begin{tabular}{|p{0.12\linewidth}|p{0.42\linewidth}|p{0.42\linewidth}|}
\hline
\makecell{Property} & \makecell{Evidence from Metal-Poor Halo Stars} & \makecell{Evidence from Dwarf Galaxies} \\ \hhline{|=|=|=|}
Rate & 
\makecell{Rare: large [Eu/Fe] scatter} & 
\makecell{Rare: fraction of ultra-faint dwarf galaxies \\ 
with $r$-process; \\
${\sim}10^{-3}$ that of core-collapse supernovae} \\ \hline
Yield & 
\makecell{Ambiguous due to unknown dilution mass \\ and metal-poor star formation site} & 
\makecell{At least one prolific $r$-process site \\ $M_r \gtrsim 10^{-2} M_\odot$} \\ \hline
Delay Time & 
\makecell{Likely have prompt sources \\ and may have delayed sources} & 
\makecell{Must have both prompt and delayed sources} \\ \hline
Composition & 
\makecell{Universal pattern from 2nd to 3rd peak \\ Variations in relative abundance \\of 1st peak and actinides} & 
\makecell{Consistent with halo stars} \\ \hline
\end{tabular}
\caption{\label{summarytable} Summary of Evidence for $R$-process Site(s)}
\end{center}
\end{table}

In more detail:
\begin{itemize}

\item There are large variations in the absolute level of observed $r$-process enhancements between stars, indicating either variable yields, or that the gas mass in which yields were diluted were very different, or both. 

\item There are significant variations in the detailed elemental composition of $r$-process-enhanced stars. There is a universal pattern from the 2nd to 3rd $r$-process peak between the solar system $r$-process residuals, but the relative amount of first peak and actinide elements changes from star to star. Future observations should be able to quantitatively investigate these variations in metal-poor halo stars.

\item Some low-mass dwarf galaxies (e.g. Reticulum\,II) show $r$-process enhancement while most others do not, indicating early $r$-process enrichment is stochastic. Thus, the observed large $r$-process scatter in metal-poor halo stars is likely due to accreting dwarf galaxies of different masses, rather than inhomogeneous scatter within a single source of metal-poor stars. 

\item In higher-mass dwarf galaxies, very few (if any) of the most metal-poor stars appear to be completely free of $r$-process material, indicating that some $r$-process element enrichment is rare but not delayed, i.e. it is ubiquitous after a certain amount of stellar mass has been achieved in any given system.
However, the most metal-rich stars in these dwarf galaxies have differing amounts of $r$-process relative to core-collapse supernovae ([Eu/Mg]), showing that delayed sources eventually dominate.

\item Taken together, $r$-process dwarf galaxies and metal-poor $r$-process halo stars point to a requirement for both prompt and delayed sources of $r$-process elements.

\end{itemize}

\subsection{New Surveys and New Dwarf Galaxies}

The majority of results so far have come from studying the Milky Way stellar halo and intact dwarf galaxies. The next decade will bring a number of exciting spectroscopic and deep photometric surveys that will push this field to new levels. Over the coming years, the community can thus anticipate the discovery of many more dwarf galaxies whose varying properties will reveal more information about the nature of the $r$-process. \\

\textit{New intact dwarf galaxies to be found in large photometric surveys:}
Given current sky coverage, it is unlikely (though not impossible) that many new intact dwarf galaxies will be found around the Milky Way with $M_\star \gtrsim 10^6 M_\odot$ in the coming years.
However, it is extremely likely that over 100 ultra-faint dwarf galaxies with $M_\star \lesssim 10^5 M_\odot$ will be found \citep{Hargis2014,Manwadkar2022}, especially with the Vera Rubin Observatory's Legacy Survey of Space and Time \citep{Ivezic2019}. With spectroscopic followup, this should lead to a much better mapping of and understanding about early stochastic $r$-process enrichment.\\

\textit{Tidally disrupted galaxies discovered with Gaia and spectroscopic surveys:}
In principle, metal-poor stars in the halo all come from tidally disrupted dwarf galaxies, and it is starting to be possible to associate them with erstwhile dwarf galaxies.
Likely only  the surface has been scratched of discovering tidally disrupted dwarf galaxy streams and halo structures with \emph{Gaia} and spectroscopic surveys. Hence, the chemical study of such systems is still in relative infancy \citep{Roederer10,Ji2020b,Hansen2021,Aguado2021,Matsuno2021,Matsuno2022}.
In the near future, major spectroscopic surveys with the capability to directly measure $r$-process abundances are WEAVE and 4MOST \citep{WEAVE,FOURMOST}. The surveys SDSS-V and DESI are lower spectral resolution and do not have the capability to measure $r$-process elements, but they will help discover structures that can then be followed up with other facilities.
There are many nascent plans to build other wide-field spectroscopic facilities that could potentially benefit $r$-process studies in both intact and tidally disrupted dwarf galaxies, such as the Mauna Kea Spectroscopic Explorer (MSE; \citealt{TheMSEScienceTeam2019}).
These tidally disrupted dwarf galaxies should help fill out the range of dwarf galaxy masses and star formation durations (Figure~\ref{fig:dwarfprop}), enabling testing different chemical evolution predictions on a larger sample of $r$-process abundance measurements.
\\

\subsection{New Instrumentation and Observational and Experimental Facilities}

\textit{Ultraviolet spectroscopy:}
Many of the most important $r$-process elements can only be measured at ultraviolet wavelengths, but it is currently difficult to study even the brightest most nearby stars \citep{Roederer2012,Roederer2022} this way. The United States 2020 Astronomy Decadal Survey identified a large IR-O-UV space telescope as a priority for future development, and a high-resolution UV spectrograph would help expand on existing, very limited stellar UV datasets. However, it is unlikely that even future UV spectrographs will have enough light-collecting power to study stars in distant dwarf galaxies.\\

\textit{Infrared spectroscopy:}
Other than ultra-violet, one technological development in the past years is the development of high quality infrared detectors and high-resolution spectrographs. The promise of this for $r$-process studies is not yet clear because absorption lines in the infrared tend to be weaker than the optical. However, lines have been identified for Sr, Y, Zr, Ba, Sm, Ce, Nd, Eu, Dy, and Yb \citep{Matsunaga2020,Smith2021}) in metal-rich, solar-type disk stars.
\\

\textit{Extremely Large Telescopes:}
The vast majority of spectroscopic observations of dwarf galaxies are currently done on telescopes with effective mirror diameters of 6.5\,m to 10\,m (e.g., the \textit{Magellan} telescopes, the Very Large Telescopes, the Subaru Telescope, the Keck Telescopes).
A major endeavor of the next decade is three projects aiming to build telescopes with effective mirror diameters of $25-40$\,meters: the Giant Magellan Telescope (GMT), the Thirty Meter Telescope (TMT), and the European Extremely Large Telescope (E-ELT).
Tripling the telescope diameter effectively triples the distance out to which dwarf galaxies can be studied while also enabling much larger numbers of stars to be studied in individual galaxies.
It also enables much higher signal-to-noise spectra of existing stars, which will likely enable the detection of additional elements, especially thorium.\\

\textit{New Experimental Facilities:} Currently, theoretical calculations of $r$-process nucleosynthesis rely on extensive models of nuclear masses and reactions. The nuclear data uncertainties significantly limit many predictions of astrophysical $r$-process calculations \citep[e.g.,][]{Eichler2015,Mendoza-Temis2015,Mumpower2016,Surman2017,Barnes2021}.
However, upcoming experimental facilities like the Facility for Rare Isotope Beams (FRIB), the Radioactive Isotope Beam Factor (RIBF), the International Facility for Antiproton and Ion Research (FAIR), the Advanced Rare IsotopE Laboratory (ARIEL), and the $N=126$ Factory at Argonne will fundamentally transform this landscape \citep{Horowitz2019}. In the future, small elemental and isotopic abundance deviations currently deemed consistent will likely become major areas of research.
\\\\

The next decades thus promise important advances in the field of nuclear astrophysics, especially the study of the $r$-process and its astrophysical site. 
Combining results from metal-poor $r$-process enhanced stars in the Galactic halo and dwarf galaxies with other observations, such as gravitational wave events from neutron star mergers, deep sea radioactive isotope measurements of recent, local $r$-process events, as well as experimental measurements of nuclei and reaction rates and theoretical predictions for supernovae and neutron star merger yields has the potential to provide a unified picture of the origin of $r$-process elements across cosmic time.

\section*{Acknowledgments}
We thank our colleagues in the field of nuclear astrophysics for many years of fruitful and enjoyable discussions about $r$-process topics, both observationally and theoretically, and seeing the field grow from individual stellar abundance patterns to large samples, including observations of stars in dwarf galaxies. 
Specifically, we warmly thank our collaborators
Tim Beers,
Rana Ezzeddine,
Terese Hansen,
Erika Holmbeck,
Evan Kirby,
Ting Li,
Gail McLaughlin,
Andy McWilliam,
Rohan Naidu,
Brian O'Shea,
Vinicius Placco,
Ian Roederer,
Charli Sakari,
Joshua Simon,
and
Rebecca Surman
for sharing our excitement about nuclear astrophysics. 
We are grateful to the Joint Institute for Nuclear Astrophysics (JINA) for facilitating many of these interactions over the last 20 years.
We acknowledge support awarded by the U.S. National Science Foundation (NSF): AST 1716251 and PHY 14-30152; Physics Frontier Center/JINA Center for the
Evolution of the Elements (JINA-CEE).


\end{document}